\documentclass[10pt,final]{IEEEtran}
\usepackage{amsmath}
\usepackage{amsfonts}
\usepackage{epsfig}
\usepackage{amssymb}
\usepackage{cite}
\usepackage{hhline}
\usepackage{multirow}
\usepackage{algorithmic}
\usepackage[ruled]{algorithm}
\usepackage{color}
\usepackage{bbm}

\begin{document}

\title{Analysis and Optimization of Random Sensing Order in Cognitive Radio Networks}

\author{Hossein Shokri-Ghadikolaei and Carlo Fischione %
\thanks{Manuscript received Jan 5, 2014; revised April May 11, 2014 and Aug 7, 2014.
This work was supported by the FP7 EU project Hydrobionets. This paper was presented in part at the IEEE International Conference on Communications (ICC), Sydney, Australia, June 2014.}
\thanks{The authors are with KTH Royal Institute of Technology, Stockholm, Sweden (e-mail: \{hshokri and carlofi\}@kth.se).}}

\maketitle

\begin{abstract}
  Developing an efficient spectrum access policy enables cognitive radios to dramatically increase spectrum utilization while ensuring predetermined quality of service levels for primary users. In this paper, modeling, performance analysis, and optimization of a distributed secondary network with random sensing order policy are studied. Specifically, the secondary users create a random order of available channels upon primary users return, and then find optimal transmission and handoff opportunities in a distributed manner. By a Markov chain analysis, the average throughputs of the secondary users and average interference level among the secondary and primary users are investigated. A maximization of the secondary network performance in terms of the throughput while keeping under control the average interference is proposed. It is shown that despite of traditional view, non-zero false alarm in the channel sensing can increase channel utilization, especially in a dense secondary network where the contention is too high.
  Then, two simple and practical adaptive algorithms are established to optimize the network. The second algorithm follows the variations of the wireless channels in non-stationary conditions and outperforms even static brute force optimization, while demanding few computations. The convergence of the distributed algorithms are theoretically investigated based on the analytical performance indicators established by the Markov chain analysis.
  Finally, numerical results validate the analytical derivations and demonstrate the efficiency of the proposed schemes. It is concluded that fully distributed sensing order algorithms can lead to substantial performance improvements in cognitive radio networks without the need of centralized management or message passing among the users.
\end{abstract}


\section{Introduction}\label{section: Intro}
\PARstart{E}{merging} new wireless applications and ever-growing
demands for a higher data rate challenges the limited spectrum resources and consequently the current fixed spectrum allocation policies. To effectively mitigate the problems associated with the fixed
spectrum allocation, the promising concept of cognitive radio networks (CRNs) has been the focus of intense research in both academic and regulatory bodies~\cite{wang2011emerging,wang2011advances}.

CRN promotes spectrum utilization by allowing low priority secondary users (SUs) to opportunistically exploit the under-utilized licensed channels of high priority primary users (PUs) in an intelligent manner~\cite{Gavrilovska2013learning}.
Meanwhile, due to preemptive priority of the
PUs to access the channels, the SUs must vacate the channel whenever
the corresponding PUs appear. In this case, a set of procedures called spectrum
handoff (SHO) is initiated to help the SUs in finding new transmission opportunities, through reliable spectrum sensing, and resuming their unfinished transmissions~\cite{christian2012spectrum}. Clearly, the performance of an SHO framework depends heavily on the performance of spectrum sensing. The noise and the channel impairments such as shadowing and fading, however, lead to decision errors, quantified in terms of false alarm and misdetection probabilities. A false alarm occurs when a free channel is mistakenly sensed busy, while a miss
detection happens whenever an occupied channel is sensed free.
With each false alarm, a transmission opportunity is lost, and after each misdetection, an SU starts to transmit on the channel and consequently interferes with signal present in the channel.

\subsection{Spectrum Handoff Models}\label{section: handoff model}
Broadly speaking, SHO procedures can be modeled by connection-based and slot-based modeling techniques~\cite{wang2012modeling}. The connection-based model defines the spectrum handoff upon appearance of the PUs (event-driven manner), while in the slot-based methods, the spectrum handoff process can be performed in each time slot (time-driven manner). In~\cite{wang2012modeling,LCWang2010CM,wang2012optimal2012,zhang2013what}, performance of connection-based SHO in terms of extended data delivery time and handoff delay is extensively evaluated, and several optimization framework for SHO are proposed. In this paper, we focus on the slot-based SHO model.

An SU can conduct wideband or narrowband spectrum sensing at the beginning of each time slot, depending on the power budget and affordable computational complexity. In the wideband spectrum sensing, an SU senses multiple channels simultaneously, while only one channel can be sensed at a time in the narrowband spectrum sensing~\cite{sun2013wideband}. Easier implementation, lower power consumption, and less computational complexity lead to great interest in narrowband spectrum sensing. Here, we assume that the SUs are able to sense and possibly transmit on one channel at a time. In this case, an SU sorts the channels in an order, called sensing order, and transmits on first channel that is sensed free in the established order.
If the channel is sensed busy, the SU initiates the SHO procedure and then senses the second channel of the sensing order, and so on. Such a sensing-access is called sequential channel sensing~\cite{christian2012spectrum}.

\begin{table*}[]
\centering
  \caption{Comparison of the existing solutions for spectrum handoff and sequential channel sensing. LSES stands for limited sensing in each slot. Yes/No means that the paper considers throughput in each iteration of its algorithm, however does not find overall throughput, considering the impact of other users' transmissions.}\label{table: related_work}
{\renewcommand{\tabcolsep}{3pt}
   \begin{tabular}{c c c c c c c c c c c c c c c c c c c c c c c c}
\hline \hline
   & Number of SUs & Distributed & False alarm & Misdetection & Throughput & Intra-CRN interference & Inter-CRN interference & Sequential channel sensing & LSES \\ \hline
\cite{zhang2013what} & Multiple & \checkmark & \checkmark & -- & -- & -- & -- & \checkmark & -- \\ \hline
\cite{shokriSMS} & Multiple & -- & \checkmark & \checkmark & \checkmark & -- & -- & \checkmark & \checkmark \\ \hline
\cite{pei2011energy} & Single & -- & \checkmark & \checkmark & Yes/No & -- & -- & \checkmark & -- \\ \hline
\cite{Eryigit2013Energy} & Multiple & -- & \checkmark & \checkmark & -- & -- & -- & \checkmark & -- \\ \hline
\cite{Zheng2011target} & Two & -- & -- & -- & -- & -- & -- & \checkmark & -- \\ \hline
\cite{cheng2011simple} & Multiple & \checkmark & -- & -- & Yes/No & -- & -- & \checkmark & -- \\ \hline
\cite{Li2013Almost} & Multiple & \checkmark & -- & -- & Yes/No & -- & -- & \checkmark & \checkmark \\ \hline
\cite{Zhang2013Channel} & Single & -- & \checkmark & \checkmark & Yes/No & -- & -- & \checkmark & -- \\ \hline
\cite{Poor} & Single & -- & \checkmark & \checkmark & \checkmark & -- & -- & \checkmark & \checkmark \\ \hline
\cite{zhang2012cognitive} & Single & -- & \checkmark & \checkmark & \checkmark & -- & -- & \checkmark & \checkmark \\ \hline
\cite{Fan2009} & Two & -- & -- & -- & Yes/No & -- & -- & \checkmark & -- \\ \hline
\cite{Liu2013Traffic} & Multiple & -- & -- & -- & \checkmark & -- & -- & \checkmark & \checkmark \\ \hline
\cite{Misra2012optimal} & Multiple & -- & -- & -- & \checkmark & -- & -- & \checkmark & -- \\ \hline
\cite{shokriISMS} & Multiple & -- & \checkmark & \checkmark & \checkmark & -- & -- & \checkmark & \checkmark \\ \hline
\cite{Shokri2014Analysis} & Multiple & -- & \checkmark & \checkmark & \checkmark & -- & \checkmark & \checkmark & \checkmark \\ \hline
\cite{Kim2013Optimal} & Multiple & -- & \checkmark & \checkmark & Yes/No & -- & -- & \checkmark & \checkmark \\ \hline
\cite{ewaisha2011optimization} & Single & -- & \checkmark & \checkmark & \checkmark & -- & \checkmark & \checkmark & -- \\ \hline
\cite{Paysarvi2012On} & Single & -- & \checkmark & \checkmark & \checkmark & -- & \checkmark & \checkmark & -- \\ \hline
\cite{song2012prospect} & Multiple & \checkmark & -- & -- & -- & -- & -- & -- & -- \\ \hline
\cite{khan2013Autonomous} & Multiple & \checkmark & \checkmark & -- & -- & -- & -- & \checkmark & -- \\ \hline
\cite{ShokriTWireless} & Multiple & \checkmark & \checkmark & \checkmark & \checkmark & -- & -- & \checkmark & \checkmark \\ \hline
  \hline
   Ours & Multiple & \checkmark & \checkmark & \checkmark & \checkmark & \checkmark & \checkmark & \checkmark & \checkmark \\ \hline

  \hline
\end{tabular}}
\end{table*}

\subsection{Related Work}\label{section: related work}
Recently, the problem of designing a proper framework for sequential channel sensing has gained much interests. In~\cite{pei2011energy}, the optimal sensing order design has been investigated in order for an SU to achieve the maximum energy efficiency by applying a dynamic programming solution. The tradeoff between sensing accuracy and consumed energy in sequential channel sensing is investigated in~\cite{Eryigit2013Energy}, wherein optimal solution along with two suboptimal heuristic algorithms are proposed for determining proper sensing time and order that maximize the energy efficiency. In~\cite{Zheng2011target}, the authors find the optimal sensing order to minimize the probability of not finding a free channel upon triggering of SHO.
Besides energy efficiency, throughput maximization is also extensively studied.
Optimal and suboptimal sensing orders of a CRN containing only one SU are developed in~\cite{cheng2011simple,Li2013Almost,Zhang2013Channel,Poor,zhang2012cognitive}, which maximize the average achievable throughput of the SU in a time slot. These results have been further extended for a CRN with two~\cite{Fan2009} and multiple SUs~\cite{Liu2013Traffic,Misra2012optimal,shokriSMS,Shokri2014Analysis,shokriISMS}. A closed-form optimal solution as well as three suboptimal solutions for maximizing the average throughput by setting proper sensing orders have been proposed in~\cite{shokriSMS}, and an intelligent sensing order setting scheme for a centralized CRN has been introduced in~\cite{shokriISMS}.

In~\cite{Kim2013Optimal}, a dynamic programming-based framework for sequential channel sensing is proposed to minimize the SHO delay for a heterogeneous network.
Although maximizing the average throughput is of critical importance in secondary communications design, the final framework might be of difficult applicability, since throughput maximization may result in a large interference with the primary network, which violates the prerequisites of a CRN, being transparent to the primary network. Therefore, a general framework considering both throughput and interference is desirable. In~\cite{ewaisha2011optimization}, the authors investigate the optimal sensing time and order for maximizing the expected throughput of a CRN with one SU, and for penalizing interferences that disrupt the primary communications.
The same problem is investigated in~\cite{Paysarvi2012On}, wherein optimal parameters for spectrum sensing, i.e, the sensing time and decision threshold, are found.

\subsection{Motivation}\label{section: motivation}
Most of the literatures in the slot-based SHO focus on single SU or centralized CRNs~\cite{pei2011energy,Liu2013Traffic,Poor,zhang2012cognitive,Fan2009,shokriSMS,shokriISMS,Shokri2014Analysis,Kim2013Optimal,Misra2012optimal,ewaisha2011optimization,Paysarvi2012On}, where the existence of a coordinator is an inseparable part of these centralized algorithms. The coordinator computes the best parameters for the optimum networks operation, and then let the SUs know the parameters. The main problem is that not only a centralized network coordinator cannot be assumed in many CRNs’ applications, but it imposes a massive computational burden on the network as well. As shown in~\cite{shokriSMS}, for instance, the computational complexity of finding the optimal sensing orders exponentially increases with number of PUs and number of SUs. Actually, this holds even in single SU case~\cite{Li2013Almost}, which motivates the authors to develop a suboptimal SHO algorithm.

An SHO framework for a distributed CRN without a common control channel is proposed in~\cite{song2012prospect}.
However, a wideband and perfect spectrum sensing is considered in the paper, which are hard to be applicable in many practical systems.
In~\cite{khan2013Autonomous}, an autonomous weighting policy is developed with the aim of minimizing the likelihood of collisions with other SUs in a distributed manner.
The authors show that their algorithm might achieve collision-free sensing orders, i.e., the SUs never collide among them. However, the misdetection probability is assumed zero, meaning that the SUs would not make interference for the PUs as well as other SUs. Therefore, quality of service (QoS) provisioning for the PUs is not addressed.
In~\cite{ShokriTWireless}, the authors exploit a modified p-persistent MAC protocol to set the sensing orders of the SUs in a distributed manner. However, it is assumed that the SUs can successfully transmit on the channel even if the PU presets on the channel. In fact, they focused on maximum achievable throughput and did not study the interference (inter or intra) in the network. As a result, there is no QoS guaranteeing mechanism for the PUs. Also, the authors assumed that the spectrum sensing performance does not change, even though the level of signals present in the channel changes. Table~\ref{table: related_work} summarizes the strengths and weaknesses of the main representative approaches in sequential channel sensing. The last column of the table shows if an SU can sense only limited number of potential channels in a time slot.

\subsection{Contribution}\label{section: contribution}
In this paper, we substantially extend our previous study~\cite{ShokriTWireless} and extend the investigations of~\cite{Paysarvi2012On} to a multiuser distributed CRN, where not only the interference among the SUs and PUs, hereafter called inter-CRN interference, is important, but also the interference among the secondary users, hereafter called intra-CRN interference, is highly important, since it affects the QoS of the secondary connections. We investigate the performance of a CRN adopting random sensing order policy. That is, once an SHO is triggered, all the SU create a set of random channels to be sensed, and sequential channel sensing process is initiated.
Although sequential channel sensing with random sensing order policy is not a new problem, and it has been used as a baseline scheme to make comparison, e.g.,~\cite{Li2013Almost}, the analytic performance evaluation and optimization of such a prominent scheme are not well addressed in the literature.
We propose a finite state novel Markov chain to effectively model the SUs, which enables us to capture
the interactions of the users in the network and find average throughput, intra- and inter-CRN interferences, collision and successful transmission probabilities, average number of handoffs, average
delay for the SUs before starting a transmission, and many other important performance measures. These results are established without making several unrealistic simplifications, for example assuming zero miss-detection probability~\cite{zhang2013what, Zheng2011target, cheng2011simple,Li2013Almost, Misra2012optimal, song2012prospect,khan2013Autonomous}, that almost ignores the existence of interferences in the network. We derive the average throughput of the SUs, intra- and inter-CRN interferences, as the main performance metrics, and then formulate an optimization problem to maximize the average throughput while keeping the average interferences bounded.

This optimization, however, poses a high computational burden, and it is not always consistent with real non-stationary wireless channels. Therefore, we propose novel cross-layer adaptive and distributed algorithms of light computational complexity that use no analytical models for the link statistics, where the brute force solution of the optimization problem is used as a benchmark to check the performance of the distributed algorithms.
By these algorithms, each SU just needs to receive ACKs from its receiver to iteratively maximize the average throughput for a given maximum allowable inference. Furthermore, as the proposed algorithms can be implemented in a fully distributed fashion without any need of message passing among the SUs, they decrease the problems associated with control channel establishment in CRN terminology~\cite{LiangChen,Lo}.
We prove that the proposed algorithms converge to the solution of the centralized optimization problem in expectation, in probability, and almost surely, and then we study for which conditions the algorithm's parameters ensure convergence. Motivated by high performance, it is concluded that we can trust simple algorithms with minimum signaling overheads to optimize the performance of sensing order. These results challenge the need of having complex scenarios for designing optimal sensing order, including huge computational complexity~\cite{Li2013Almost, shokriSMS}, having a central coordinator~\cite{pei2011energy,Liu2013Traffic,Poor,zhang2012cognitive,Fan2009,shokriSMS,shokriISMS, Kim2013Optimal,Misra2012optimal,ewaisha2011optimization,Paysarvi2012On}, or massive signaling overhead~\cite{shokriISMS}. The proposed schemes provide a new baseline for lower bound of the performance of a distributed/centralized solution, since it is possible to achieve higher performance by imposing more signaling overheads (simply negotiating with the neighbors).
In addition, we show that the traditional view of false alarm, i.e., smaller false alarm higher average throughput, may no longer valid in a distributed CRN, where the contention among the users plays an important role. In fact, higher false alarm can substantially increase channel utilization, and this improvement is more prominent in dense secondary network scenario\footnote{A secondary network is called dense when the number of SUs, which exist in the transmission range of each other, are much higher than the number of primary channels.}, where the contention is too high, and
false alarm will contribute in collision reduction and consequently overall throughput enhancement.

Compared to the literature mentioned above, this is the first paper to 1) consider the problem of SHO for sequential channel sensing in a distributed set-up with more realistic assumptions including misdetection and false alarm probabilities, 2) investigate the inter- and intra- CRN interferences and keep both of them under control, 3) investigate the impact of the SUs' transmissions on the channel occupation probabilities and spectrum sensing performance, 4) propose simple and practical algorithms to keep the overall system performance at the optimum level while maintaining the QoS guarantees in non-stationary conditions.

The rest of this paper is organized as follows. In Section \ref{section: System Model}, we describe the considered CR network. In Section \ref{section: RSOP}, the structure of the random sensing order policy is described, and its performance is evaluated. Moreover, two efficient algorithms are proposed in Section~\ref{section: Optimization} to optimize the performance of the network.
Numerical results are then provided in Section \ref{section: Numerical Results}, followed by concluding
remarks provided in Section \ref{section: Conclusion}.

\section{System Model}\label{section: System Model}
\begin{figure}[t]
\centering
  \includegraphics[width= 8.5cm]{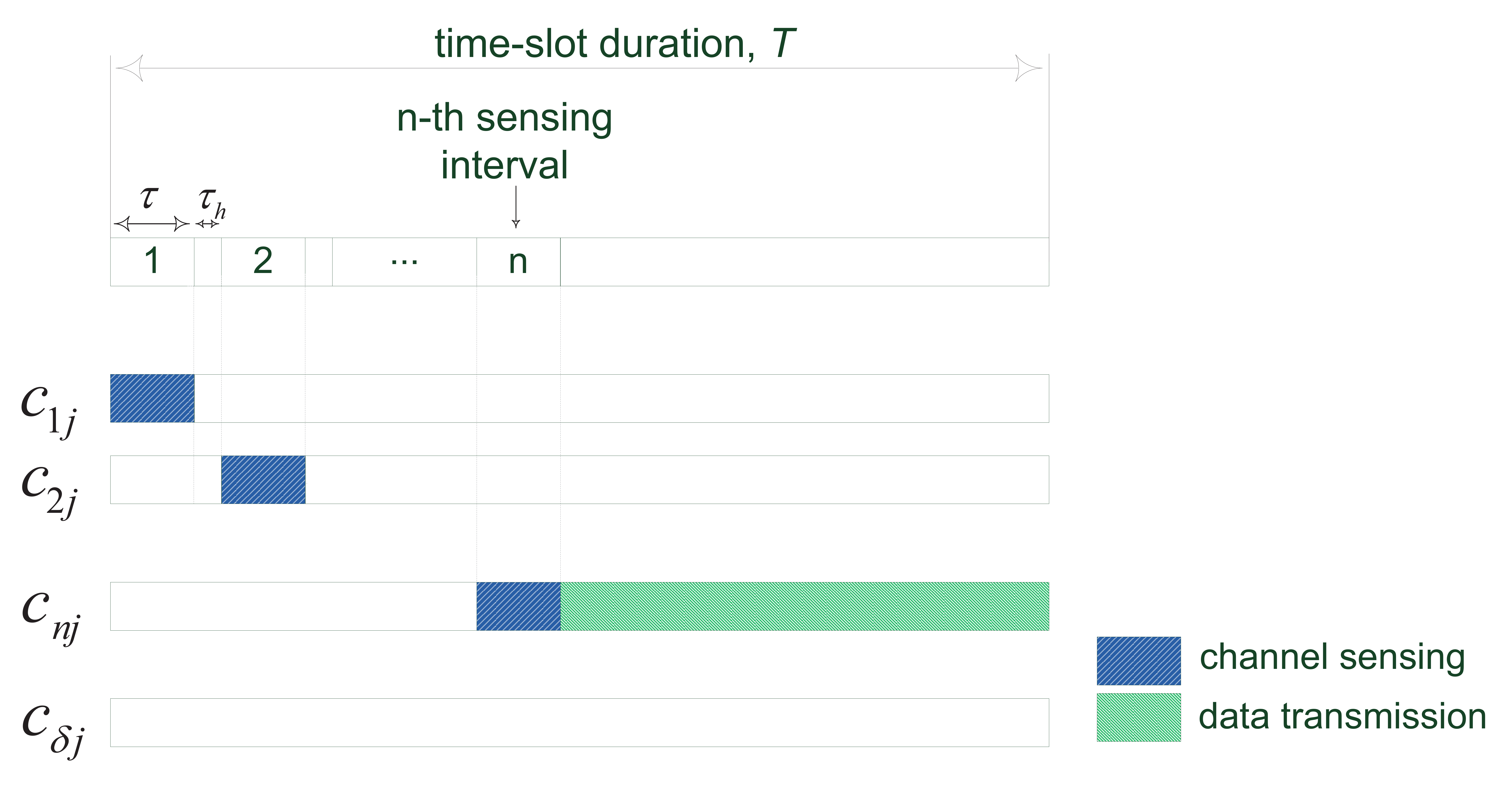}\\
  \caption{General timing structure adopted by each SU $j$.}
  \label{fig: TimingStructure}
\end{figure}

A time slotted CRN with $N_s$ SUs is assumed. The SUs attempt to
opportunistically transmit on the channels exclusively dedicated to
the $N_p$ PUs, each having one channel. As assumed in~\cite{Poor,zhang2012cognitive,Fan2009}, the SUs are synchronous in time-slots with other SUs as well as the PUs. In the sequential channel sensing methodology, once a handoff is requested, each SU's time slot divides into sensing and transmission modes. In the sensing mode, the
SUs sequentially sense the channels based on their sensing orders~\cite{Poor,zhang2012cognitive,Fan2009}.
Suppose that the sensing order of SU $n$, for $n=1,\ldots, N_s$ (see Table~\ref{table: notations} for a summary of the frequently used notations), is
\begin{equation}\label{eq2}
{c_n} = \left[ c_{1n}, c_{2n}, \ldots,c_{\delta n} \right] \:,
\end{equation}
where $c_{1n}$ and $c_{\delta n}$ denote the first and the last channels
to be sensed. $\delta$ is the maximum number of channels that an SU can sense in a time slot.
The SUs sense the first channel of their sensing orders, $c_{1,j}$ for $1\leq j \leq N_s$, and start their communications on the channels sensed free\footnote{Clearly, these transmissions might lead to collisions among the SUs or interferences with the PUs or other SUs.}. Other SUs initiate
the handoff process, which takes ${\tau}_{h}$ seconds\footnote{The
SU spends this time to change the circuitry in order to sense the next channel of its sensing order. This time duration do not depend on the amount of frequency shift
required by the reconfiguration~\cite{shokriSMS}.}, and then sense the second
channel of their sensing orders. The procedure continues until~\cite{shokriLB}: a) all the SUs find transmission opportunities, b) no time remains for sensing new channels in
the time slot, or c) no non-sensed channels remains. It holds~\cite{shokriLB} that
\begin{equation}\label{eq3}
\delta = 1 + \min \left( { \left\lfloor \frac{ {T - \tau }}{{\tau  + {\tau
_{h}}}} \right\rfloor ,{N_p} - 1} \right) \:,
\end{equation}
where $T$ is a time slot duration, and $\tau$ is the channel sensing time.
After sensing $ \left( n - 1 \right) $ occupied channels, if an SU finds $n$-th channel of its sensing order free, the user will transmit data on that channel for the rest of the slot. In the case, the time length left in the slot for the transmission is
\begin{equation}\label{eq1}
{\rm{RT}}_n = T - \tau - \left( n - 1 \right) \left( \tau + \tau_{h} \right) \:.
\end{equation}
Fig.~\ref{fig: TimingStructure} demonstrates the timing structure of each SU $j$.

\begin{table}
  \centering
  \caption{Summary of main notations}\label{table: notations}
{\renewcommand{\arraystretch}{1.1}
   \begin{tabular}{|l| l| l| l|}
\hline
   Symbol & Definition \\ \hline
    $p$ & Channel sensing probability \\ \hline
    $r$ & Average throughput of each SU \\ \hline
    $t_I$ & Average interference time in the network\\ \hline
    $C_R$ & Transmission rate \\ \hline
    $\mathcal{L}^{\left( n \right)}$ & Number of SUs sensing channel $m$ at stage $n$ \\ \hline
    $N_p$ & Number of PUs \\ \hline
    $N_s$ & Number of SUs \\ \hline
    $N_{ep}$ & Number of slots in a frame (estimation period)  \\ \hline
    $P_{{\rm fa},m}^{\left( n \right)}$ & False alarm probability of sensing channel $m$ in sensing stage $n$ \\ \hline
    $P_{d,m}^{\left( n \right)}$ & Detection probability of sensing channel $m$ in sensing stage $n$ \\ \hline
    $P_{m,1}$ & Presence probability of the PU $m$ \\ \hline
    $P_{m,1}^{\left( n \right)}$ & Probability that channel $m$ is busy at the beginning of stage $n$ \\ \hline
    $P_{{\rm md}}^{\max }$ & Maximum allowable misdetection probability \\ \hline
    $P_{{\rm fa}}^{\max}$ & Maximum allowable false alarm probability \\ \hline
    $\rm{RT}_n$ & Transmission time if $n$-th channel of sensing order is sensed free \\ \hline
    $T$ & Time slot duration \\ \hline
    $\delta$ & Maximum number of channels can be sensed in a time slot \\ \hline
    $\tau$ & Sensing time \\ \hline
\end{tabular}}
\end{table}

\section{Modeling and Performance Evaluation}\label{section: RSOP}
In this section, the random sensing order policy (RSOP) is modeled, and the performance of a CRN with RSOP is derived using a Markov chain analysis. Then, an adaptive protocol is proposed to optimize the performance of the CRN.
Recall that each SU sequentially senses the channels based on an order. It has been shown that regardless the computational complexity, the sensing orders can be optimally determined in a single user~\cite{Poor} or centralized multiple user~\cite{shokriSMS} CRNs. However, we cannot directly apply those proposals to distributed CRNs. For such networks, simple sensing orders are proposed in~\cite{ShokriTWireless}, wherein the channels are arranged by their indices. For a simple order, we have~\cite{ShokriTWireless}:
\begin{equation}\label{eq4}
c_{1j} = 1 \quad  c_{2j} = 2 \quad \ldots \quad c_{\delta j} = \delta \;, 1 \leq j \leq N_s \:.
\end{equation}
While this order facilitates the network modeling and performance evaluation, it causes a high level of contention to access the same channels, which significantly degrades the average throughput of the CRN.
Also, more efficient sensing orders, as proposed in~\cite{song2012prospect,khan2013Autonomous}, are highly sensitive to false alarm and misdetection probabilities. In fact, they are originally developed for perfect spectrum sensing, which is not achievable in real world.

In order to mitigate the aforementioned problem, we propose to use optimal RSOP. In this scheme, an SU chooses by a random distribution a target channel in each sensing interval between $1$ and $N_p$ for each $c_{ij}$. Therefore, the requests of the SUs are uniformly distributed among all available channels, and thereby the CRN throughput increases by the reduction of the contention for accessing the same channels. Moreover, adopting RSOP desirably bypasses message passing or other signaling overheads required for designing optimal sensing orders in a distributed CRN. Altogether, RSOP is highly desired if its performance is optimized. 

\begin{figure} [t]
\centering
\hspace{-0.4cm}
  \includegraphics[width= 9.2cm]{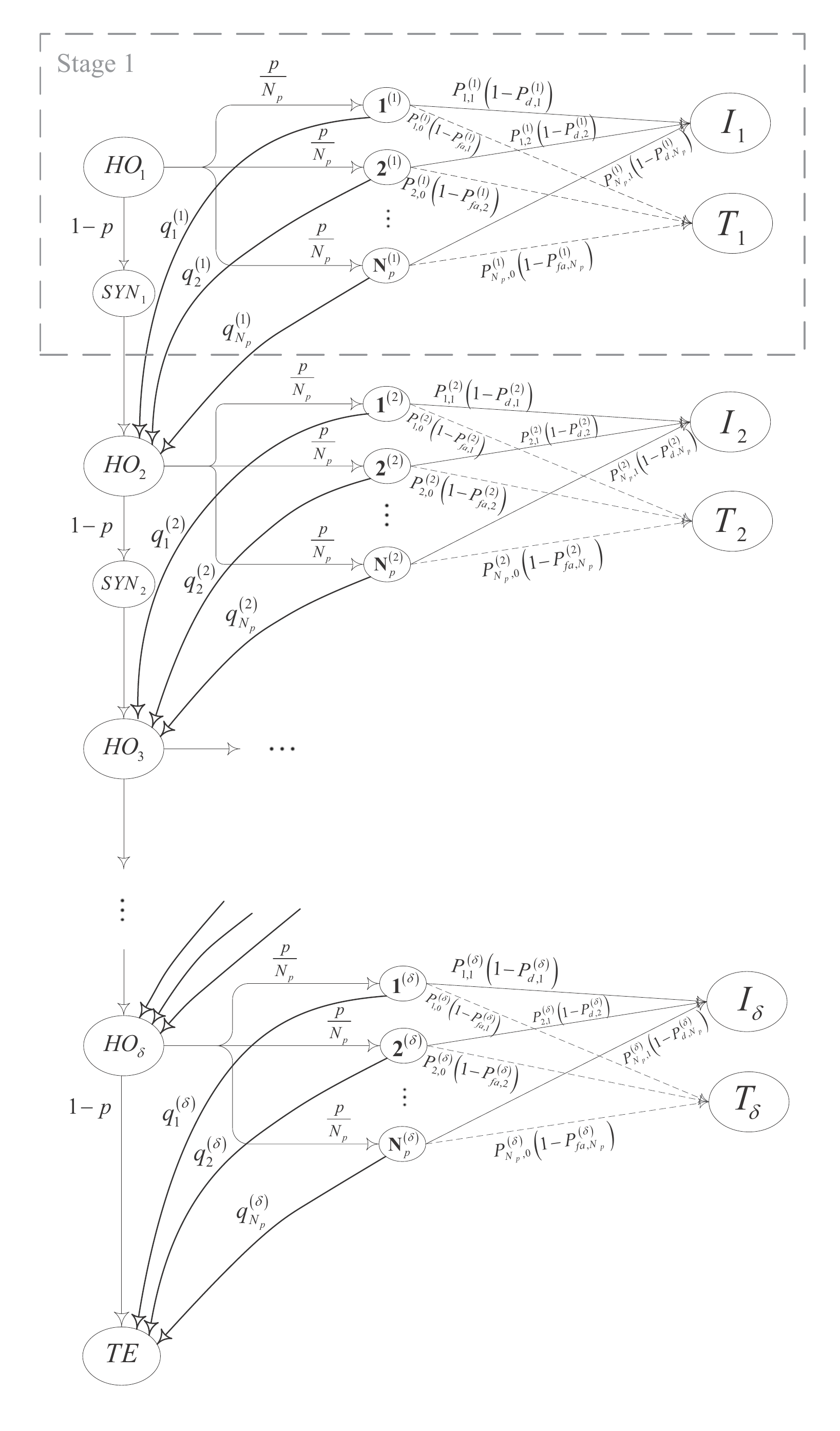}\\
  \caption{Proposed Markov chain to model the SU's behavior. The states correspond to different stages of spectrum sensing and packet transmission.}
  \label{fig: MarkovChain}
\end{figure}

We utilize a modified version of the conventional p-persistent multiple access protocol to further decrease the contention and enhance the performance of multiple access among the SUs. That is, each SU senses each channel with the probability $p$ and skips the sensing process with the probability $ \left( 1 - p \right) $.
Channel sensing probability $p$, provides a degree of freedom to optimize the performance of the CRN in the form of maximizing the average throughput and keeping the interference bounded. Such a modified p-persistent access outperforms conventional p-persistent multiple access, wherein channels are sensed at first, and then the SU decide to transmit, if found free, with probability $p$~\cite{Wang2012Energ}.
The main problem is that the channels are always sensed in the conventional p-persistent algorithms~\cite{Wang2012Energ}, while for the modified version presents here, it is possible that the SU skips this channel and starts the sensing process of the next channel (if exists). As a consequence, more channels will be sensed in the conventional p-persistent multiple access algorithm compared to the modified one.
Further comparisons between their average throughput and energy consumptions (for spectrum sensing) will present in Section~\ref{section: Numerical Results}.

Fig.~\ref{fig: MarkovChain} models by a Markov chain the channel sensing-access policy
of the RSOP used by each SU.
The state ${{m}}^{\left( n \right)}$ refers to the case that the SU starts to sense, which takes $\tau$ time units, and possibly transmits on the channel $m$ at the $n$-th sensing stage. Let $P_{m,0}^{\left( n \right)}$ and $P_{m,1}^{\left( n \right)}$ respectively be the probability that the channel $m$ is free and occupied at the beginning of the $n$-th sensing stage. Let $P_{{\rm fa},m}^{\left( n \right)}$ and $P_{{\rm md},m}^{\left( n \right)} = 1 - P_{d,m}^{\left( n \right)}$ denote the false alarm and misdetection probabilities of sensing the channel $m$ in the $n$-th sensing stage.
An SU can successfully transmit on the channel $m$ if it is free, and the false alarm does not occur. Once this event happens, with the probability $P_{m,0}^{\left( n \right)}\left( {1 - P_{{\rm fa},m}^{\left( n \right)}} \right)$, the SU's state changes to the transmitter states $T_n$, and it transmits on the channel for the rest of the time slot, ${\rm{RT}}_n$, with the constant rate of $C_R$.
Even though we consider constant transmission rate, we can easily extend the formulations present in the paper to cover heterogenous SUs. This assumption is done also in~\cite{Fan2009,Misra2012optimal}. The interference experienced at stage $n$ is denoted by $I_n$ and happens whenever the channel is busy, and the SU mistakenly senses it free, with probability $P_{m,1}^{\left( n \right)}\left( {1 - P_{d,m}^{\left( n \right)}} \right)$.
Different handoffs are modeled using states ${{\rm HO}_i}~$, $i = 1, 2, \ldots, \delta$, where recall that $\delta$ is defined in \eqref{eq3}. Note that the first handoff state does not exist in the search process, and we use it for easing the analysis without loss of generality.

At the beginning of each time slot, an SU gives the state ${\rm HO}_1$ (in Fig.~\ref{fig: MarkovChain}), and immediately transits to one of the first sensing states, ${{1}}^{\left( 1 \right)}$, ${{2}}^{\left( n \right)}$, or ${{N_p}}^{\left( 1 \right)}$ with the identical probabilities of $ {p}/{N_p} $, or to the synchronizer state ${\rm{SYN}}1$\footnote{Note that the SUs that are not routed to the sensing states, with probability $\left( 1 - p \right)$, enter standby mode (at state ${\rm{SYN}}_1$) and wait for $\tau$ time units (sensing period). Then, they are directed to the state ${\rm HO}_{2}$. With the help of the synchronizer states, all SUs will enter the $i$-th stage state at the same time.}. After $\tau$ time units, the SU's state changes to the transmitter state ${T}_1$, interference state ${I}_1$, or to the second handoff state ${\rm HO}_2$ with the probability $q_{m}^{\left( 1 \right)}$. From Fig.~\ref{fig: MarkovChain},
\begin{equation}\label{eq5}
{q_{m}^{\left( 1 \right)}} = P_{m,0}^{\left( 1 \right)}{P_{{\rm fa},m}^{\left( 1 \right)}} + P_{m,1}^{\left( 1 \right)}{P_{d,m}^{\left( 1 \right)}} \:.
\end{equation}
This procedure continues until the maximum number of admissible
handoff is reached. Let us define the $i$-th stage of the sensing-access process, shown in Fig.~\ref{fig: MarkovChain}, as the set of states of the Markov chain ${\rm HO}_i$, ${\rm{SYN}}_i$, ${{m}}^{\left( i \right)}$,
${T}_i$, and ${I}_i$. After the stage $\delta$, the SU's state transits to the terminate state ${\rm TE}$, meaning that the SU sleeps for the rest of the time slot, $T - \tau - \left( \delta - 1 \right) \left( \tau + \tau_{h} \right)$. Then, it repeats the search-access process at the beginning of the next slot~\cite{Poor}.
In the RSOP, a busy channel can be occupied either by the corresponding PU or other SUs; other SUs might detect the channel as a transmission opportunity at the previous stages. This event allows us to establish two results. First, the channel occupancies status changes in successive stages. Second, the average signal level that is present in the wireless media changes by each sensing stage. In other words, as it is possible that some SUs transmit on occupied channels in each stage $n$, the remained SUs face a higher received signal levels if they sense those channels at the stage $n+1$.

\emph{Proposition I:} Consider the Markov chain of Fig.~\ref{fig: MarkovChain}. The occupation probability of channel $m$ at the beginning of stage $n$ is
\begin{equation}\label{eq6}
P_{m,1}^{\left( n \right)} = P_{m,1}^{\left( n - 1 \right)} + {P_{m,0}}{ \left( P_{{\rm fa},m}^{\left( 1 \right)} \right) ^{\mathcal{L}^{\left( 1 \right)} + \mathcal{L}^{\left( 2 \right)} + \cdots + \mathcal{L}^{\left( n - 2\right)}}}{{U}_m ^{\left( n - 1 \right)}} \:,
\end{equation}
for $1 \leq m \leq N_p$ and $1 \leq n \leq \delta$, where $\mathcal{L}^{\left( n \right)}$ denotes the average number of SUs sensing channel $m$ at stage $n$, $U_m ^{\left( n \right)}$ is the probability of transmission on channel $m$ at stage $n$ by at least one SU conditioned on the absence of the corresponding PU, and $P_{m,0}$ denote the absence probability of the PU $m$.

\emph{Proof}: A proof is given in Appendix A. \hspace*{\fill}{$\blacksquare$}

The received SNR affects the performance of spectrum sensing and sequential channel sensing. To increase the accuracy of the RSOP model, the different detection and false alarm probabilities have to be considered in various sensing stages. The performance analysis of various spectrum sensing techniques are out of the scope of this paper. However, we derive the formulations for energy detector-based spectrum sensing and use it
for optimization purposes and numerical results.

\emph{Proposition II:} Consider the Markov chain of Fig.~\ref{fig: MarkovChain}. For the energy detector-based spectrum sensing, it holds
\begin{equation}\label{eqPfa}
P_{{\rm fa},m}^{\left( n \right)} = P_{{\rm fa},m}^{\left( 1 \right)},  \qquad 1 \leq m \leq N_p, \quad 1 \leq n \leq \delta \:,
\end{equation}
and
\begin{equation}\label{eqPmd}
P_{d,m}^{\left( n \right)} \approx P_{d,m}^{\left( 2 \right)}, \qquad 1 \leq m \leq N_p, \quad 3 \leq n \leq \delta \:,
\end{equation}
where $P_{{\rm fa},m}^{\left( 1 \right)}$, $P_{d,m}^{\left( 1 \right)}$, and $P_{d,m}^{\left( 2 \right)}$ are given in \eqref{eq204}-\eqref{eq207}.

\emph{Proof}: A proof is given in Appendix B. \hspace*{\fill}{$\blacksquare$}

\emph{Proposition III:} Consider the Markov chain of Fig.~\ref{fig: MarkovChain}. Let $T$ be the slot duration, ${Q}_{T_n,m}$ be the probability of successful transmissions of each SU at each channel $m$ from state $T_n$, ${\rm{RT}}_n$ be the remained time of the current slot, and $C_R$ be the constant transmission rate of the SUs. Let ${Z}_{I_n,m}$ be the probability that no SU cause interference on the channel $m$ at the stage $n$. Then, the average throughput of each SU is
\begin{equation}\label{eq7}
r \left( \tau , p \right) = \frac{1}{T} \sum\limits_{{m}=1}^{N_p} {\sum\limits_{n=1}^{\delta} {{Q}_{T_n,m} {\rm{RT}}_n C_R}} \:.
\end{equation}
The average interference time due to the each SU's transmissions is
\begin{equation}\label{eq8}
t_I \left( \tau , p \right) = \frac{1}{TN_p} {\sum\limits_{m = 1} ^{N_p} {\sum\limits_{n = 1} ^{\delta} { \left( 1 - {Z}_{I_n,m} \right) {\rm{RT}}_n}}} \:.
\end{equation}

\emph{Proof}: A proof is given in Appendix C. \hspace*{\fill}{$\blacksquare$}

Now that we have derived the key performance indicators, we can turn our attention to the optimal selection of the parameters that maximizes the throughput.
\section{Distributed Channel Sensing Optimization}\label{section: Optimization}
In this section, first we investigate the optimal theoretical value that the sensing time and channel sensing probability should assume. Then, we present practical distributed algorithms of light computational requirements to achieve such an optimum with an adequate accuracy.
\subsection{Theoretical Optimal Parameter Selection}
As can be observed from the propositions established in the previous section, performance measures given by \eqref{eq7} and \eqref{eq8} depend on $\tau$ and $p$, where recall that $\tau$ is the channel sensing time and $p$ is the channel sensing probability. Hence, the performance of the CRN can be maximized by optimally choosing the values of $p$ and $\tau$ that maximize the average throughput, as a QoS metric for the SU, and bounding the interference time, as a QoS metric for the PUs as well as the SUs. That is,
\begin{align} \label{eq9}
&\hspace{-0.51cm} \mathop {\text{maximize}}\limits_{\tau , p} \hspace{0.8cm} r \left( \tau , p \right) \\
&\hspace{-0.51cm} {\text{subject to}} \hspace{0.5cm} t_I \left( \tau , p \right) \leq t_I ^{\max} \:, \tag{11.a} \label{C1_Const} \\
& \hspace{1.45cm} P_{{\rm md},m}^{\left( n \right)} \left( \tau , p \right) \leq P_{{\rm md}}^{\max}, \hspace{0.3cm} 1 \leq m \leq N_p \:,  \nonumber\\
& \hspace{5cm} 1 \leq n \leq \delta  \:, \tag{11.b} \label{C2_Const} \\
& \hspace{1.5cm} 0 \leq \tau \leq T  \:, \tag{11.c} \label{C3_Const} \\
& \hspace{1.5cm} 0 \leq p \leq 1 \:, \tag{11.d} \label{C4_Const}
\end{align}
where $t_{I}^{\max}$ represents the maximum tolerable value of the average interference time, which depends on the QoS level guaranteed for the PUs as well as the SU. $P_{{\rm md}}^{\max}$ is the maximum tolerable misdetection probability imposed by the standard~\cite{StandardDarft}. Constraint \eqref{C1_Const} guarantees a QoS level for both the PUs and SUs. Constraint \eqref{C2_Const} further provides QoS for just PUs, and \eqref{C3_Const} and \eqref{C4_Const} sets the admissible values for decision variables.

By direct inspection, we see that the optimization problem is generally non-convex, making it difficult to
be efficiently solved. Such a complexity is exacerbated by that the parameter $\delta$ in the cost and constraint functions makes them not differentiable (see \eqref{eq3}).
In order to find an approximate optimal solution to \eqref{eq9}, we could search for a convex approximation of the problem~\cite{boyd2004convex}.
However, this is difficult, because it is no obvious how to find a good convex approximation and, moreover, there is no guarantee on the distance between the optimal solution and the approximated one~\cite{boyd2004convex}.

From another perspective, the dimension of decision variables is $2$ ($\tau$ and $p$), regardless the size of the networks, i.e., number of PUs and SUs. Feasible solution(s) is in a bounded box due to~\eqref{C3_Const} and~\eqref{C4_Const}.
Altogether, running a centralized brute force optimization is more reasonable than finding
an approximate sub-optimal solution. This is particularly motivated by that the availability of the optimal solution is herein interesting as a benchmark for an approximate solution provided by distributed algorithms. Indeed, one of the core contribution of this paper is to develop a distributed solution algorithm of low computational complexity that allows us to reach the optimal solution of \eqref{eq9}. Therefore, it is not essential to establish a centralized solution method of \eqref{eq9}. However, note that the results of the propositions and the formulation of the optimization problem are of paramount importance to establish the optimally of the distributed solution method, as we propose in the following.

Note that one assumption we adopted, as widely done in the literature~\cite{Misra2012optimal,shokriISMS,shokriSMS,ShokriTWireless,Kim2013Optimal,song2012prospect,khan2013Autonomous}, is that the SUs use the same channel sensing time and probability. In real world scenarios, this may not be the case because every SU may experience different wireless channel conditions.
To mitigate this issue, we develop novel fully adaptive and distributed algorithms to let the SUs follow the variation of the environment and keep the performance of the network at a near-optimal point.
In the following, we characterize such a distributed algorithm that gives the solution of \eqref{eq9}, when (a) each SU is able to adaptively change its channels sensing time and probability in each sensing stage and (b) the parameters describing the channels (e.g., PUs' traffic pattern and fading properties) change.

An interesting aspect of the cost and constraint functions that appear in \eqref{eq9} is that we can estimate them locally at each SU by taking local measurements.
This is identical to radio power control algorithms~\cite{Fischione2011Fast}, and allows developing a distributed solution for
scenario of equal channel sensing and time probabilities and the general scenario of unequal channel sensing and time probabilities. In the general scenario, the SUs may use different sensing-access parameters, depending on their own preferences.

\emph{Remark I:} The functions of optimization problem \eqref{eq9} are convex on the region of decision variables of interest, as shown by extensive Monte Carlo simulations, which we use due to the formidable complexity of an analytical investigation of the convexity. Monte Carlo simulations are common to use for establishing convexity of functions when their difficult analytical structure or nonlinearities does not allow explicit derivation of the derivatives~\cite{boyd2004convex}.

In the general scenario, an optimization problem as \eqref{eq9} can be formulated. We do not characterize analytically the cost and constraints of this general case due the analytical intractability.
Rather, using Remark I, we develop a distributed algorithm to find the optimal solution of the optimization problem \eqref{eq9}, based on stochastic subgradient method. We show, by mathematical analysis and numerical simulations, that the algorithm works well. We note that the analysis that gives the functions of problem \eqref{eq9} and the subsequent Remark I, are essential to derive the distributed algorithm.

\subsection{Distributed Sequential Channel Sensing Algorithms}
As noted in~\cite{Park06}, each SU is able to estimate the average throughput and interference time for a given $\tau$ and $p$. Also, $P_{{\rm md},m}^{\left( n \right)}$ can be calculated (see Appendix~B) by individual SU. Coherently, we can develop an algorithm considering the impact of decision variables on the cost and constraint functions.
The starting point of such algorithm is the following observation on the optimization decision variables $p$ and $\tau$:
increasing $p$ leads to higher demands for transmission (pros) and contention level in the network (cons). Reducing $\tau$, from another perspective, increases transmission time (pros) at the expense of higher false alarm and misdetection probabilities (cons).
Therefore, an SU decides for increasing/reducing $p$ and $\tau$ in each estimation period ($N_{ep}$ consecutive slots, defined as a frame) so that the average throughput increases while \eqref{C1_Const}-\eqref{C4_Const} are met. Otherwise, it adjusts decision variables in the reverse direction. This update process mimics the stochastic subgradient method, which can provide optimal solution of an optimization problem after some iterations based on noisy measurements~\cite{boyd2008stochastic}. We are now ready to present the details of the algorithm.

Let $\tau_{m}^{k}$ and $p_{m}^{k}$ denote the channel sensing time and access probability of SU $m$ at time (frame) $k$. 
$\tau_{\min}$ is the minimum value of sensing time.
Let ${\widetilde r}_m^{~k}$ and ${\widetilde {t_I}}_m^{k}$ be the unbiased estimated throughput and interference at SU $m$ based on $\tau_{m}^{k}$ and $p_{m}^{k}$.
Given that optimization problem \eqref{eq9} is convex (recall Remark I), we can develop an iterative distributed algorithm in which every SU updates its optimization parameters locally using ${\widetilde r}_m^{~k}$ and ${\widetilde {t_I}}_m^{k}$. Then, they operate with the updated optimization parameters $\tau_{m}^{k+1}$ and $p_{m}^{k+1}$, which certainly affect average throughput and interference. Once a reduction in the average throughput is detected (${\widetilde r}_m^{~k} < {\widetilde r}_m^{~k-1}$) by an increase of the channel sensing time ($\tau_{m}^{k} > \tau_{m}^{k-1}$), the transmitter should change the update direction and adopt a smaller value for the next channel sensing time, i.e., $\tau_{m}^{k+1} < \tau_{m}^{k}$, in order to increase the transmission time and consequently the throughput. Otherwise, the transmitter is in the right direction of increasing the channel sensing time.
From another perspective, if the reduction in the channel sensing time ($\tau_{m}^{k} < \tau_{m}^{k-1}$) leads to a throughput loss (${\widetilde r}_m^{~k} < {\widetilde r}_m^{~k-1}$), the transmitter should enhance the quality of spectrum sensing by adopting a higher value for the next channel sensing time, i.e., $\tau_{m}^{k+1} > \tau_{m}^{k}$ to compensate the observed throughput loss. Otherwise, the transmitter can still decrease the channel sensing time and thereby increase transmission time without devastating the quality of the spectrum sensing (see Fig.~\ref{fig: Avg Thr and Int ver tau} and its discussions).
Similar claims can be used for updating the channel sensing probability. Note that the adjusted values of $\tau_{m}^{k+1}$ and $p_{m}^{k+1}$ are then projected onto $\left[0,T \right]$ and $\left[0,1\right]$ to met \eqref{C3_Const} and \eqref{C4_Const}, respectively.
In a nutshell, ${\widetilde r}_m^{~k}$, ${\widetilde {t_I}}_{m}^{k}$, and $P_{{\rm md},m}^{\left( n \right)}$ will be compared to ${\widetilde r}_m^{~k-1}$, $t_I^{\max}$, and $P_{{\rm md},m}^{\max}$, and the signs of these comparisons determine whether the adopted directions for increasing/decreasing of the decision variables are correct or should be altered. Note that \eqref{C2_Const} would be met by properly selecting a minimum value for sensing time, which discussed later. However, we design our distributed algorithm for general problem without excluding \eqref{C2_Const}.
Let ${\mathcal{A}}_{m}^{k} = \{ {\widetilde r}_m^{~k} \geq {\widetilde r}_m^{~k-1}  \} \cap  \{ {\widetilde {t_I}}_m^{k} \leq t_{I}^{\max} \} \cap \{P_{{\rm md},m}^{\left( n \right)} \leq P_{{\rm md},m}^{\max}\}$ define the event of having a higher estimated throughput and meeting interference and misdetection constraints.
We define by ${\mathcal{B}}_{m}^{k} = \{ \tau_{m}^{k} \geq \tau_{m}^{k-1}\}$ and ${\mathcal{C}}_{m}^{k} = \{ p_{m}^{k} \geq p_{m}^{k-1}\}$ the events of increasing the channel sensing time and access probability, respectively. Let ${\mathcal{D}}_{m}^{k} = {\text{XOR}} \left( {\mathcal{A}}_{m}^{k} , {\mathcal{B}}_{m}^{k} \right)$ be the exclusive OR operation over ${\mathcal{A}}_{m}^{k}$ and ${\mathcal{B}}_{m}^{k} $, taking true if they do not have the same value. Let ${\mathcal{E}}_{m}^{k} = {\text{XOR}} \left( {\mathcal{A}}_{m}^{k} , {\mathcal{C}}_{m}^{k} \right)$, and $\alpha_k \geq 0$ be the step size at time $k$. The update procedure is the following:
\begin{equation} \label{eq: convergence1}
\tau_{m}^{k+1} = \left[ \tau_{m}^{k} - \mathbbm{1}_{{\mathcal{D}}_{m}^{k}} {\Delta\tau} \alpha_k  \right]^\dag \:,
\end{equation}
and
\begin{equation} \label{eq: convergence2}
p_{m}^{k+1} = \left[ p_{m}^{k} - \mathbbm{1}_{{\mathcal{E}}_{m}^{k}} {\Delta p} \alpha_k  \right]^\ddag \:,
\end{equation}
where $\left[\cdot\right]^\dag$ and $\left[\cdot\right]^\ddag$ represent the simple projection operation of sensing time and access probability onto box $\left[0,T\right]$ and $\left[0,1\right]$, respectively, and
\begin{equation} \label{eq: convergence3}
\mathbbm{1}_{\centerdot} = \left\{ {\begin{array}{*{20}{l}}
  {+1}&{{\text{if}} ~ {\centerdot} ~ {\text{is true}}} \\
  {-1}&{\text{otherwise}}
\end{array}} \right. \:.
\end{equation}
The proposed distributed algorithm is summarized in Algorithm~\ref{Algorithm: Alg1}.

{
\renewcommand{\baselinestretch}{1.05}
\begin{algorithm} [t]
\caption{\small Distributed sequential channel sensing algorithm for SU $m$}
\label{Algorithm: Alg1}
\begin{algorithmic}[1]
\small
\STATE \textbf{Initialization:} Choose $\tau_{\min}$ and initial values for ${\tau}_{m}^{1}$ and ${p}_{m}^{1}$. Set
$\rm{counter} \leftarrow 0 $, ${\widetilde r}_m^{~1} \leftarrow 0$, ${\tau}_{m}^{0} \leftarrow \tau_{\min} $, and ${p}_{m}^{0} \leftarrow 0 $.
\FOR{each slot}
\IF{$ {\rm{counter}} = {N_{ep}}$}
\STATE Calculate and then report ${\widetilde r}_m^{~k}$, ${\widetilde {t_I}}_m^{k}$, and $P_{{\rm md},m}^{\left( n \right)}$.
\STATE Compute $\mathbbm{1}_{{\mathcal{D}}_{m}^{k}}$ and $\mathbbm{1}_{{\mathcal{E}}_{m}^{k}}$.
\STATE $\tau_{m}^{k+1} \leftarrow \left[ \tau_{m}^{k} - \mathbbm{1}_{{\mathcal{D}}_{m}^{k}} {\Delta\tau} \alpha_k  \right]^\dag$
\STATE $p_{m}^{k+1} \leftarrow \left[ p_{m}^{k} - \mathbbm{1}_{{\mathcal{E}}_{m}^{k}} {\Delta p} \alpha_k  \right]^\ddag $
\STATE $k \leftarrow k + 1$.
\STATE $\rm{counter} \leftarrow 0$.
\ENDIF
\STATE ${\rm{counter}} \leftarrow {\rm{counter}} + 1$.
\ENDFOR
\end{algorithmic}
\end{algorithm}
}
{
\renewcommand{\baselinestretch}{1.05}
\begin{algorithm}
\caption{\small Distributed Adaptive sequential channel sensing algorithm for SU $m$, considering fine tuning}
\label{Algorithm: Alg2}
\begin{algorithmic}[1]
\small \STATE \textbf{Initialization:} Choose $\tau_{\min}$ and initial values for ${\tau}_{m}^{1}$, and ${p}_{m}^{1}$. $\rm{counter} \leftarrow 0 $, ${\widetilde r}_m^{~1} \leftarrow 0$, ${\tau}_{m}^{0} \leftarrow \tau_{\min} $, and ${p}_{m}^{0} \leftarrow 0 $.
\FOR{each slot}
\IF{$ {\rm{counter}} = {N_{ep}}$}
\STATE Calculate and then report ${\widetilde r}_m^{~k}$, ${\widetilde {t_I}}_m^{k}$, and $P_{{\rm md},m}^{\left( n \right)}$.
\STATE Compute $\mathbbm{1}_{{\mathcal{D}}_{m}^{k}}$ and $\mathbbm{1}_{{\mathcal{E}}_{m}^{k}}$.
\STATE $\tau_{m}^{k+1} \leftarrow \left[ \tau_{m}^{k} - \mathbbm{1}_{{\mathcal{D}}_{m}^{k}} {\Delta\tau} \alpha_k  \right]^\dag$
\STATE $p_{m}^{k+1} \leftarrow \left[ p_{m}^{k} - \mathbbm{1}_{{\mathcal{E}}_{m}^{k}} {\Delta p} \alpha_k  \right]^\ddag $
\STATE $k \leftarrow k + 1$.
\STATE $\rm{counter} \leftarrow 0$.
\ENDIF
\FOR{$n=1$ to $\delta$}
\STATE $\tau_m \left[ n \right] \leftarrow \max \left( \tau_{\min} , \tau_{m}^{k} - \left( n - 1 \right) \Delta \tau _{1}\right) $
\STATE $p_m \left[ n \right] \leftarrow \min \left( 1 , p_{m}^{k} + \left( n - 1 \right) \Delta p_{1}\right)$
\ENDFOR
\STATE ${\rm{counter}} \leftarrow {\rm{counter}} + 1$.
\ENDFOR
\end{algorithmic}
\end{algorithm}
}

Though Algorithm~\ref{Algorithm: Alg1} performs well, as confirmed by numerical results that we will show later on next section, we can achieve an improved solution by further tuning the decision variables in each slot of a frame.
In the RSOP, some SUs enter the transmission or interference states at some stages like the $n$-th stage and consequently do not continue the search process among the channels. Therefore, in the average sense, less number of SUs further participate in the search process in next stage, i.e., stage $n+1$. Moreover, with a higher value for each channel sensing probability $p$, more SUs contend for accessing the channels. As a consequence, appropriately changing the channel sensing probability can lead to an increase in the achieved SUs throughput. From another perspective, it is possible that an SU experiences higher level of the energy when it senses a channel in the stage $n$ compared to the stage $n+1$. Therefore, to achieve the same sensing performance, sensing time can be decreased~\cite{liang}, and thereby the SUs will have more transmission time in the consecutive stages.
We formalize this in the next algorithm.

Let ${p}_{m} \left[ n \right]$ and ${\tau}_{m} \left[n\right]$ denote the channel sensing probability and sensing time of SU $m$ at stage $n$, respectively. We now allow the SUs to adjust the channel sensing probability as well as sensing time in each stage of a slot.
In each slot, an SU increases its sensing probabilities from $p_{m}^{k}$ to 1 at frame $k$ to increase the chance of participating in sensing-access procedures.
Similarly, sensing time will be decreased from $\tau_{m}^{k}$ to $\tau_{\min}$ to increase the time left for the transmission.
The SU $m$ starts with $p_m \left[ 1 \right] = p_{m}^{k}$ and $\tau_m \left[ 1 \right] = \tau_{m}^{k}$. Then, it linearly increases (decreases) the channel sensing probability (sensing time) in every stage. Meanwhile, the estimation and decision processes are periodically performed in each frame of $N_{\rm ep}$ slots, which update $p_{m}^{k}$ and $\tau_{m}^{k}$. Algorithm~\ref{Algorithm: Alg2} summarizes the proposed procedures. It is clear that the nonadaptive protocol is a special case of the adaptive one for $p_m \left[ n \right] = p$, $\tau_m \left[ n \right] = \tau$, and zero step size $\alpha_k = 0$ for all $m$, $n$, and $k$. Also, it reduces to Algorithm~\ref{Algorithm: Alg1} by
\begin{equation*}
\tau_m \left[ n \right] = \tau_{m}^{k} \qquad {\text{and}} \qquad {p_m \left[ n \right] = p_{m}^{k} \qquad 1 \leq n \leq \delta}
\end{equation*}
in each frame. In a nutshell, Algorithm~\ref{Algorithm: Alg1} adjusts decision variables in each frame (coarse tuning), allowing the SUs to follow the variations of the environment, whereas further adjusting of the decision variables in each slot (fine tuning) enables Algorithm~\ref{Algorithm: Alg2} to optimize the performance of the SUs in each slot as well.

In Algorithm~\ref{Algorithm: Alg2}, the initial values are set as follow. Roughly speaking, $\Delta \tau$ and $\Delta \tau _{1}$ should be small fractions of $T$, since the sensing time of each SU should be finely tuned by the algorithms considering the fact that a slight variation in the sensing time can change the performance of the spectrum sensing and thereby the average throughput and interference time substantially~\cite{Paysarvi2012On}.
Estimation period $N_{ep}$ follows a tradeoff between the estimation accuracy of the performance measures and agility of the algorithms. Larger value of $N_{ep}$, for instance, provides a more accurate estimation of the average throughput, meanwhile makes the algorithms lazy. That is, an SU cannot follow the dynamic of the environment very fast, and reaches to the optimal point slowly.
$\Delta p$ and $\Delta p_{1}$, from another perspective, regulate the contention level in the networks, and their proper values depend heavily on the size of the primary and secondary networks, i.e., the number of PUs and SUs.
Higher value of $\alpha_k$ makes some sudden changes in the value of sensing time, channel access probability, and consequently transmission rate, leading to a faster response to the congestion at the expense of higher fluctuations, even oscillations, in the average throughput.
Also, in order to find a proper value for ${\tau _{\min }}$, we should firstly recall that energy detector is used for spectrum sensing in the numerical results section. For such a spectrum sensing scheme, $\tau ^{\min}$ exists so that false alarm and misdetection probabilities become lower than a certain threshold~\cite{liang}. It holds
\begin{equation} \label{eq: minimum sensing time}
{\tau ^{\min }} = \frac{1}{{{\gamma ^2 f_s}}}{\left( {{Q^{ - 1}}\left( {P_{{\rm fa}}^{\max }} \right) - {Q^{ - 1}}\left( {P_d^{\min}} \right)\sqrt {1 + 2\gamma } } \right)^2} \:,
\end{equation}
where $f_s$ is the sampling frequency, $\gamma$ is the SNR at the SU receiver. For simplicity of presentation, we use \eqref{eq: minimum sensing time} for initializing the proposed algorithms, i.e., ${\tau _{\min }} = \tau ^{\min }$.

\subsection{Convergence Analysis}\label{sec: convergence_analysis}
In this subsection, we present the convergence analysis for Algorithm~\ref{Algorithm: Alg1}, however a similar analysis can prove the convergence and optimality of the Algorithm~\ref{Algorithm: Alg2}.
We define ${\mathbf{x}}_m^{k} \triangleq \left[ {\tau}_{m}^{k} , p_{m}^{k}\right]^T$ and ${\mathbf{\widetilde g}}_m^{k} \triangleq \left[ \mathbbm{1}_{{\mathcal{D}}_{m}^{k}} {\Delta \tau} , \mathbbm{1}_{{\mathcal{E}}_{m}^{k}} {\Delta p}\right]^T$, where $\left[ \cdot \right]^T$ is the transposition operation, thus \eqref{eq: convergence1} and \eqref{eq: convergence2} are written as
\begin{equation} \label{eq: convergence4}
{\mathbf{x}}_m^{k+1} = {\mathbf{x}}_m^{k} - {\mathbf{\widetilde g}}_m^{k} \alpha_k \:.
\end{equation}
Each SU $m$ updates its channel sensing time and access probability in each step using \eqref{eq: convergence4}.
For notation simplicity, we assumed a synchronous update of the channel sensing time and access probability. However, the actual update can be done in an asynchronous manner by the SU. The case of asynchronous updating can be easily considered with trivial, yet involved, notations.
Defining ${\mathbf{x}}^{k} \triangleq \left[ {\mathbf{x}}_1^{k} , {\mathbf{x}}_2^{k}, \ldots, {\mathbf{x}}_{N_s}^{k}\right]^T$ and ${\mathbf{\widetilde g}}^{k} \triangleq \left[ {\mathbf{\widetilde g}}_1^{k} , {\mathbf{\widetilde g}}_2^{k}, \ldots, {\mathbf{\widetilde g}}_{N_s}^{k}\right]^T$, update processes of all $N_s$ SUs can be written as
\begin{equation} \label{eq: convergence5}
{\mathbf{x}}^{k+1} = {\mathbf{x}}^{k} - {\mathbf{\widetilde g}}^{k} \alpha_k \:.
\end{equation}

We have the following proposition:

\emph{Proposition IV:} Let $f_{\text{best}}^{k} = \min \{ f({ \mathbf{x}}^{1} ),f({\mathbf{x}}^{2} ),\ldots, f({ \mathbf{x}}^{k}) \}$ be the function value for the best point found so far, where $f=-r$ is the objective function of optimization problem \eqref{eq9}. Let $f^{*} > - \infty $ be the optimum of problem \eqref{eq9}, and $\mathbb{E}$ be the expectation operation. Let $\alpha_k$ be the step size at time $k$, which are square-summable but not summable,
\begin{equation} \label{eq: convergence5-1}
\alpha_k \geq 0, \qquad \sum_{k=1}^{\infty}{{\alpha}_{k}^{2}} < \infty, \qquad \sum_{k=1}^{\infty}{{\alpha}_{k}} = \infty  \:.
\end{equation}
Let $\mathbf{x}^{*}$ be the minimizer of $f$, and the constants $G$ and $R$ satisfy $\mathbb{E} \left\| {\mathbf{\widetilde g}}^{k} \right\|_2^2 \leq G^2$ and $ \mathbb{E} \left\| {\mathbf{x}}^{1} - {\mathbf{x}}^{*} \right\|_2^2 \leq R^2$.
Then, the iterations of \eqref{eq: convergence5}, namely the iterations of Algorithm 1, converge in expectation, i.e.,
\begin{equation} \label{eq: convergence6}
\lim_{k \to \infty} \mathbb{E} ~ f_{\text{best}}^{k} \to f^{*} \:,
\end{equation}
converge in probability, i.e., for any $\varepsilon>0$
\begin{equation} \label{eq: convergence6-1}
\lim_{k \to \infty} \textbf{Prob}\left( \left| f_{\text{best}}^{k} - f^{*} \right| \geq \varepsilon \right) = 0 \:,
\end{equation}
and converges almost surely, i.e.,
\begin{equation} \label{eq: convergence6-2}
\textbf{Prob}\left( \lim_{k \to \infty}  f_{\text{best}}^{k} = f^{*} \right) = 1 \:.
\end{equation}
Moreover, the following convergence bound holds:
\begin{equation} \label{eq: convergence9}
\mathbb{E} ~ \| f_{\text{best}}^{k} - f^{*} \| \leq \frac{R^2 + G^2 \sum_{i=1}^{\infty}{\alpha_{i}^{2}}}{2\sum_{i=1}^{k}{\alpha_{i}}} \:.
\end{equation}

\emph{Proof:} The proof consists in showing that the iterations~\eqref{eq: convergence5} are the classic iterations of the stochastic subgradient method~\cite{boyd2008stochastic}. Accordingly, ${\mathbf{\widetilde g}}^{k}$ is the stochastic subgradient (see Appendix~D), and we show that such a stochastic subgradient has bounded norm. For all $k$, we have
\begin{equation} \label{eq: convergence7}
\begin{split}
\mathbb{E} \left\| {\mathbf{\widetilde g}}^{k} \right\|_2^2 &= \mathbb{E} \sum_{m=1}^{N_s} {\Delta\tau^2\left( \mathbbm{1}_{{\mathcal{D}}_{m}^{k}}\right )^2 + \Delta p^2\left( \mathbbm{1}_{{\mathcal{E}}_{m}^{k}}\right )^2} \\
&=\mathbb{E} \sum_{m=1}^{N_s} {\Delta\tau^2 + \Delta p^2} = N_s \left(\Delta\tau^2 + \Delta p^2 \right) \triangleq G^2 \:,
\end{split}
\end{equation}
were note that the expectation is taken with respect to the events ${\mathcal{D}}_{m}^{k}$ and ${\mathcal{E}}_{m}^{k}$.
Then,
\begin{equation} \label{eq: convergence8}
\mathbb{E} \left\| {\mathbf{x}}^{1} - {\mathbf{x}}^{*} \right\|_2^2 \leq N_s \left( T^2 + 1^2 \right) \triangleq R^2 \:.
\end{equation}
Note that in \eqref{eq: convergence8}, we use that $\tau_{m}^{1}$ and $p_{m}^{1}$ are in the box $\left[0,T\right]$ and $\left[0,1\right]$ for $1 \leq m \leq N_s$. From~\eqref{eq: convergence7} and~\eqref{eq: convergence8}, the convergence, as well as the bound on the optimal cost function, follow immediately by applying the proof of the convergence of the classic stochastic subgradient method in~\cite{boyd2008stochastic}.
\hspace*{\fill}{$\blacksquare$}

\emph{Corollary:} Consider Algorithm~\ref{Algorithm: Alg1}, the global iterations \eqref{eq: convergence5}, and the assumption of Proposition IV. The convergence in expectation, i.e., $\mathbb{E} ~ \|f_{\text{best}}^{k} - f^{*} \| \leq \varepsilon$ for any $\varepsilon > 0$, is ensured if the algorithm parameters $\Delta \tau$ and $\Delta p$ are chosen such that $\sum_{i=1}^{k}{\alpha_{i}\left(2 \varepsilon - {N_s \left( \Delta \tau^2 + \Delta p^2 \right) \alpha_{i}}\right)} \geq 0$.

\emph{Proof:} Consider the converge properties of the algorithm. From (22), we have
\begin{equation} \label{eq: prooofcorollary1}
\mathbb{E} ~ \|f_{\text{best}}^{k} - f^{*} \| \leq \frac{R^2 + G^2 \sum_{i=1}^{\infty}{\alpha_{i}^{2}}}{2\sum_{i=1}^{k}{\alpha_{i}}} \leq \varepsilon \:,
\end{equation}
implying that
\begin{equation} \label{eq: convergence10}
\sum_{i=1}^{k}{\alpha_{i}} \geq \frac{R^2}{2\varepsilon} + \frac{G^2}{2\varepsilon} \sum_{i=1}^{\infty}{\alpha_{i}^{2}} \:,
\end{equation}
and hence
\begin{equation} \label{eq: convergence11}
\sum_{i=1}^{k}{\alpha_{i}\left(1 - \frac{G^2\alpha_{i}}{2\varepsilon}\right)} \geq \frac{R^2}{2\varepsilon} + \frac{G^2}{2\varepsilon} \sum_{i=k+1}^{\infty}{\alpha_{i}^{2}}   \:.
\end{equation}
Since the right hand side of \eqref{eq: convergence11} is strictly greater than zero, we can draw the conclusion that
\begin{equation} \label{eq: convergence12}
\sum_{i=1}^{k}{\alpha_{i}\left(2 \varepsilon - {N_s \left( \Delta \tau^2 + \Delta p^2 \right) \alpha_{i}}\right)} \geq 0
\end{equation}
must hold, which proves the corollary.
\hspace*{\fill}{$\blacksquare$}

Further analysis on the convergence speed along with sensitivity analysis are presented in the next section.

\section{Numerical Results}\label{section: Numerical Results}
In this section, we investigate the performance of the RSOP as well as the efficiencies of the proposed adaptive protocols by simulating a network of SUs performing sequential channel sensing.
\subsection{Simulation Set-up}\label{section: simulation set-up}
To set up a simulation environment, the values of $P_{d}^{\min}$, $P_{{\rm fa}}^{\max}$, time slot duration
$T$, and the value of sampling frequency used by the energy detector, are chosen according to \emph{IEEE 802.22} standard~\cite{StandardDarft}. Table~\ref{table: Simulation parameters} summarizes the
descriptions and values of the parameters considered for the
simulations. Using a Monte Carlo simulation, the average throughput
and the average interference time are computed after simulating the
scenarios for $10^4$ times.

\begin{figure}
\centering
  \includegraphics[width= 8.5cm]{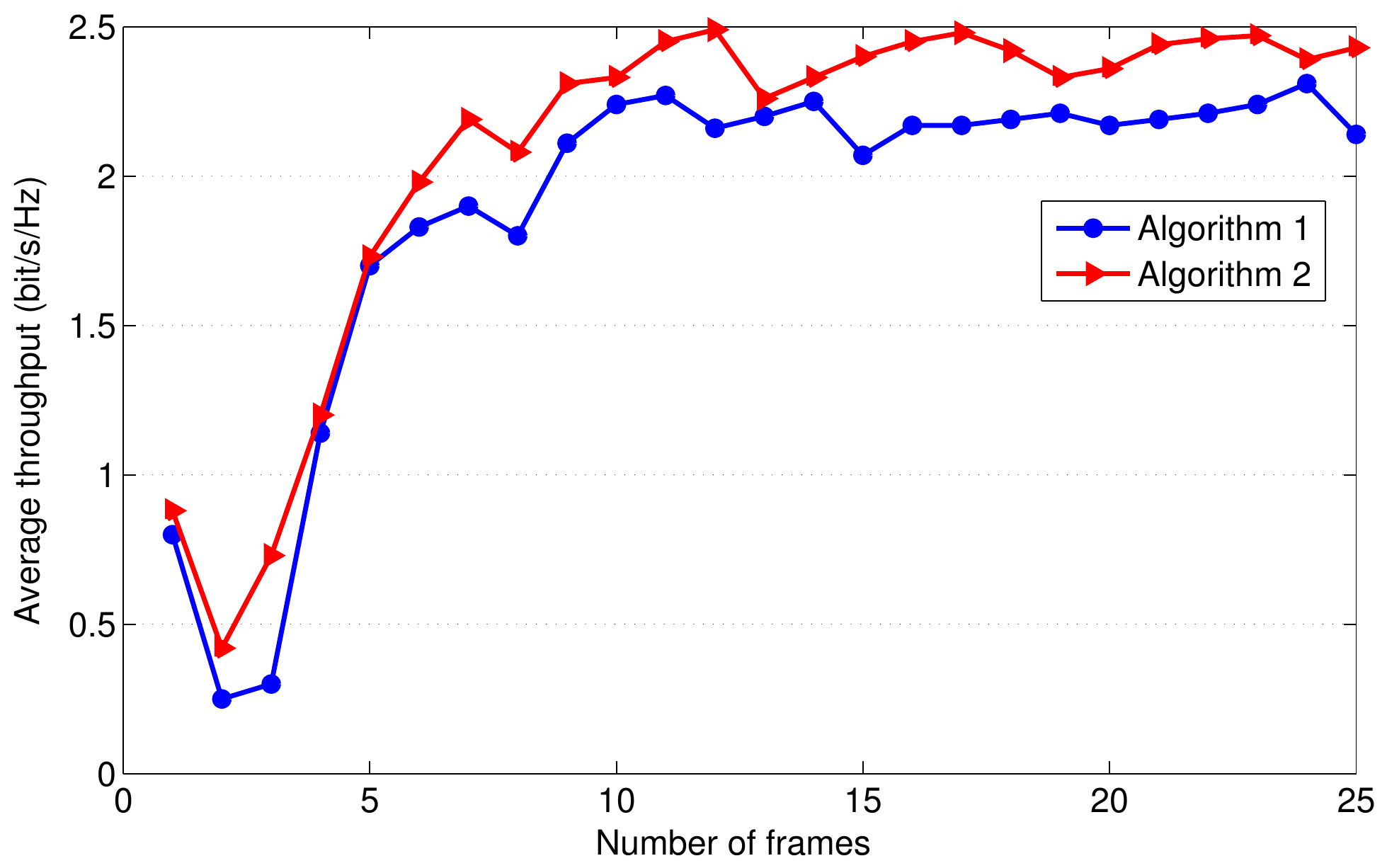}\\
  \caption{Time behaviors of the proposed algorithms ($N_s = 20$, and $N_p = 5$).}
  \label{fig: Convergence_rate}
\end{figure}

To simulate the proposed algorithms, we use $N_{ep} = 50$, $\Delta \tau = \Delta \tau _{1} = 0.01T$, $\Delta p = \Delta p_{1} = 0.025$, and $\alpha_k = 1/k$.
The utilized initial values are just an example to illustrate the effectiveness of the proposed algorithms, and one can easily investigate the impact of the aforementioned parameters on the performance of the algorithm such as convergence speed (see Section \ref{section: Optimization}), and then tries to adopt optimal initial values. However, as we proved, the algorithms will converge in all cases.
Using those parameters, the time behaviors of the algorithms are depicted in Fig.~\ref{fig: Convergence_rate}.
Note that since the proposed algorithms are based on stochastic subgradient method, the convergence time depends on each realization of the algorithms. In Fig.~\ref{fig: Convergence_rate}, we have depicted the convergence time based on our simulations as obtained by an average over 100 realizations of the proposed algorithms.

Table~\ref{table: sensitivity_analysis} analyzes the sensitivity of Algorithm~\ref{Algorithm: Alg1} to the initial parameters used in Fig.~\ref{fig: Convergence_rate}.
From the table, if we decrease $N_{\rm ep}$ by $50\%$ from $N_{\rm ep} = 50$, while keeping fixed all other parameters, Algorithm~\ref{Algorithm: Alg1} will converge after 355 iterations or equivalently after $\left(50\times0.5\right)\times355=8875$ slots. Lower $N_{\rm ep}$ leads to higher estimation error that in turn deteriorates the convergence time, while higher $N_{\rm ep}$ delays update time of the algorithm, resulting in a prolonged convergence time. For instance, 1050 and 1350 slots are required for the convergence, when $N_{\rm ep}$ increases by $8\%$ and $50\%$, respectively. Similar analyses can be conducted for other parameters.

\begin{table}
  \centering
  \caption{Simulation Parameters}\label{table: Simulation parameters}
{\renewcommand{\arraystretch}{1.1}
   \begin{tabular}{l l l l l l}
\hline \hline
   Parameter & Description & Value \\ \hline
    $T$ & Time slot duration & 10 {\text{ms}} \\
    $P_d^{\min }$ & Minimum allowable detection probability & 0.9 \\
    $P_{{\rm fa}}^{\max}$ & Maximum allowable false alarm probability & 0.1 \\
    $t_I^{\max}$ & Maximum allowable interference time & 0.05 $T$ \\
    $f_s$ & Receiver sampling frequency & 6.857 {\text{MHz}} \\
    ${\tau}_{h}$ & Required time for performing a handoff & 0.1 {\rm $\mu$s}\\
    $CR$ & Transmission rate & 1 {\text{bit/s/Hz}} \\
    $p$ & Channel sensing probability & 0.8 \\
  \hline \hline
\end{tabular}}
\end{table}

\begin{table}
\centering
\hspace{-0.5cm}
  \caption{Sensitivity analysis of Algorithm~\ref{Algorithm: Alg1}. The number of iterations is considered as the convergence time.}
{\renewcommand{\tabcolsep}{2pt}\renewcommand{\arraystretch}{1.2}
    \begin{tabular}{c|c|c|c|c|c|c|c|c|c|c|c|c|c|c|c}
    \hhline{~---------}
    \hhline{~---------}
      & \multicolumn{3}{c|}{$N_{\rm ep}$} & \multicolumn{3}{c|}{$\Delta \tau$} & \multicolumn{3}{c|}{$\Delta p$} \\
    \hhline{~---------}
      & $-50\%$ & $+5\%$ & $+50\%$ & $-50\%$ & $+5\%$ & $+50\%$ & $-50\%$ & $+5\%$ & $+50\%$ \\
    \hline
     \multicolumn{1}{|c|}{Simulation} & 355 & 20 & 18 & 29 & 16 & $18$ & 21 & $17$ & 18 \\
    \hline
    \end{tabular}}
  \label{table: sensitivity_analysis}
\end{table}

\subsection{Effects of Simulation Parameters}\label{section: effect of parameters}
\begin{figure}[t]
\centering
  \includegraphics[width= 8.5cm]{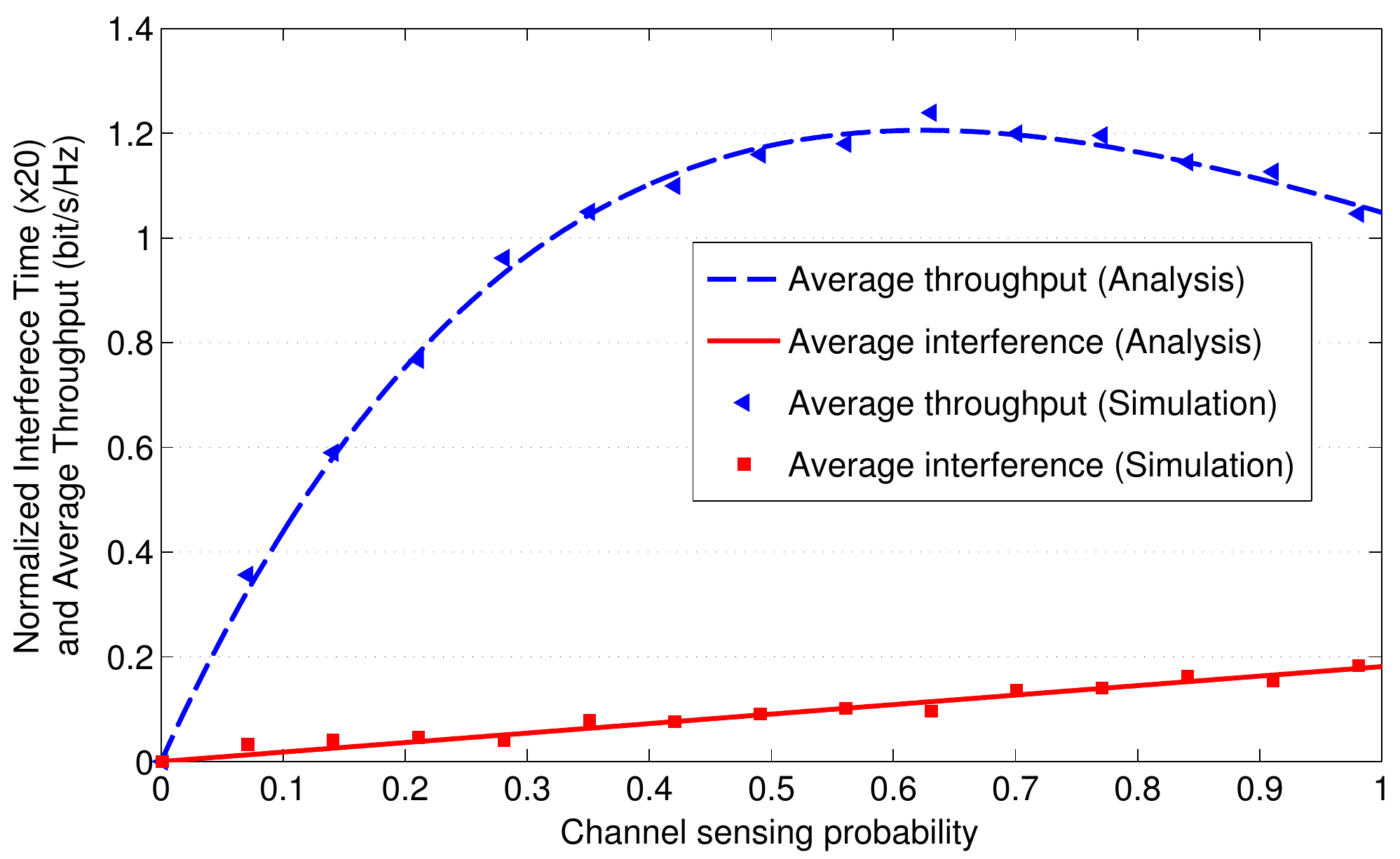}\\
  \caption{Average throughput and interference against channel sensing probability (for $\tau = 0.1 T$, $N_s = 20$, and $N_p = 5$).}
  \label{fig: Avg Thr and Int ver p}
\end{figure}

\begin{figure}[t]
\centering
  \includegraphics[width= 8.5cm]{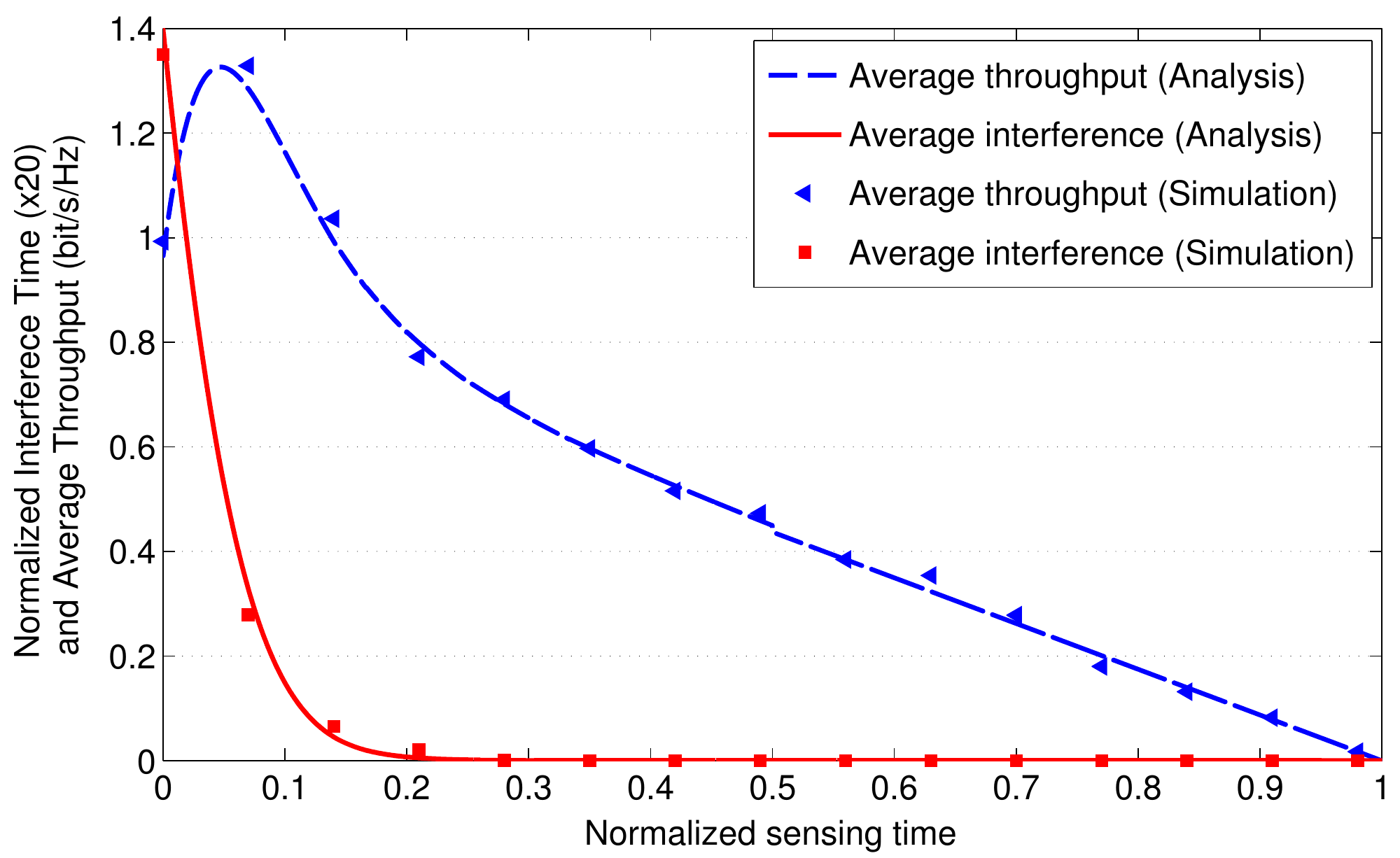}\\
  \caption{Average throughput and interference against normalized sensing time (for $N_s = 3$ and $N_p = 7$).}
  \label{fig: Avg Thr and Int ver tau}
\end{figure}

Figs. \ref{fig: Avg Thr and Int ver p} and \ref{fig: Avg Thr and Int ver tau} depict the average throughput of the secondary network and the normalized interference time, versus channel sensing probability $p$ and normalized sensing time $\tau /T$.
Clearly, as the channel sensing probability increases, the chance of finding a transmission opportunity (correctly or mistakenly) raises as well, leading to higher values for average throughput and interference. However, after an optimum point, the throughput reduces due to high connection level among the SUs.
Moreover, by incrementing the channel sensing time, the SUs sense
the channels more accurately, find more transmission opportunity,
and hence reach a higher average throughput. For the same
reason, the average interference among the SUs and PUs is
reduced. Moreover, the well known sensing-throughput tradeoff~\cite{liang} is verified. That is, after an optimum point, wherein the false alarm and misdetection
probabilities are in acceptable levels, the average throughput
starts decreasing due to the reduction of the time left for the
transmission.

\subsection{False Alarm Paradox}\label{section: false_alarm_paradox}
\begin{figure}[t]
\centering
  \includegraphics[width= 8.5cm]{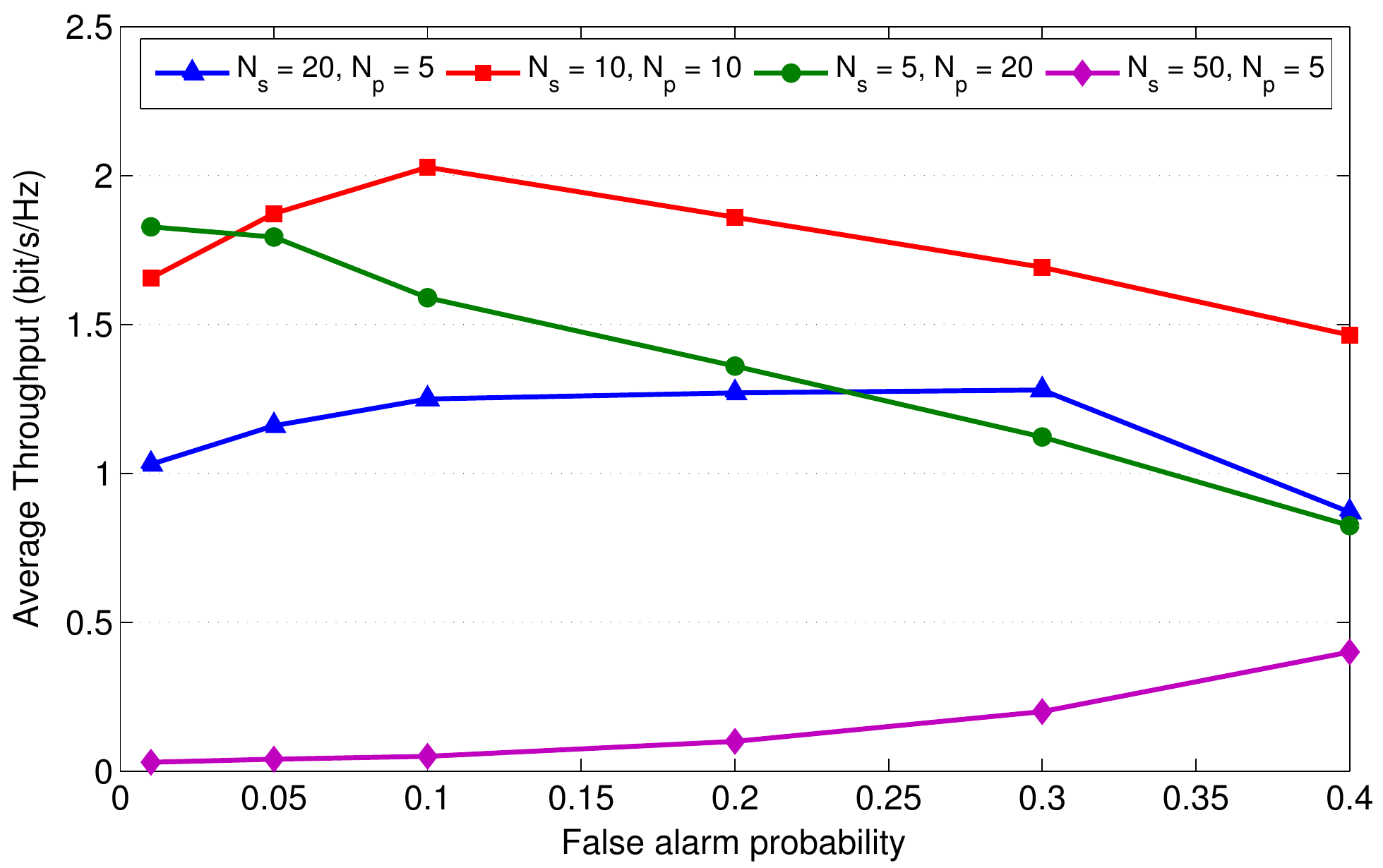}\\
  \caption{Impact of false alarm on the average throughput ($p = 0.8$).}
  \label{fig: false_alarm_effect}
\end{figure}

After a false alarm, not only an SU misses a transmission opportunity, but also less time remains for possible transmission due to the time wasted for sensing the current channel. Therefore, false alarm reduces the average transmission rate (and consequently throughput) in traditional view.
In a network with several uncoordinated users, however, increasing transmission rate of each individual user does not necessarily lead to throughput enhancement, due to possible collisions. In the case, higher false alarm probabilities might increase the total throughput by reducing the contention level of accessing the same channels, provided that it does not lead to miss major parts of the available transmission opportunities.
One can argue that false alarm's effect is the same as the effect of the channel sensing probability $p$. The main difference is that increasing $p$ will directly affect the interference level of the network (see Fig.~\ref{fig: MarkovChain}) as well. Fig.~\ref{fig: false_alarm_effect} confirms potential positive effect of false alarm on the network performance\footnote{Note that there is a well-know relation between false alarm and detection probabilities, described by receiver operating characteristic curve~\cite{Arsalan09}. In the energy detection-based spectrum sensing, however, we can change the false alarm probability for a fixed value of detection probability by adjusting sensing time and decision threshold.}. This performance improvement is more prominent in dense secondary network scenario, where the number of SUs, which exist in the transmission range of each other, are much
higher than the number of primary channels. In particular, with $N_p = 5$, adopting $P_{\rm fa}=0.3$ leads to $24\% $ and $567\% $ performance improvement compared to $P_{\rm fa}=0.01$, in a CRN with 20 and 50 SUs, respectively. Due to lack of the space, providing further analysis for finding the optimal false alarm probability for RSOP is left for future studies.

In the next subsection, we demonstrate the efficiencies of the proposed algorithms by studying the evolution of the average throughput with respect to the number of PUs and SUs.


\subsection{Effects of Network Size}\label{section: effect of netwirk size}
\begin{figure}[t]
\centering
  \includegraphics[width= 8.5cm]{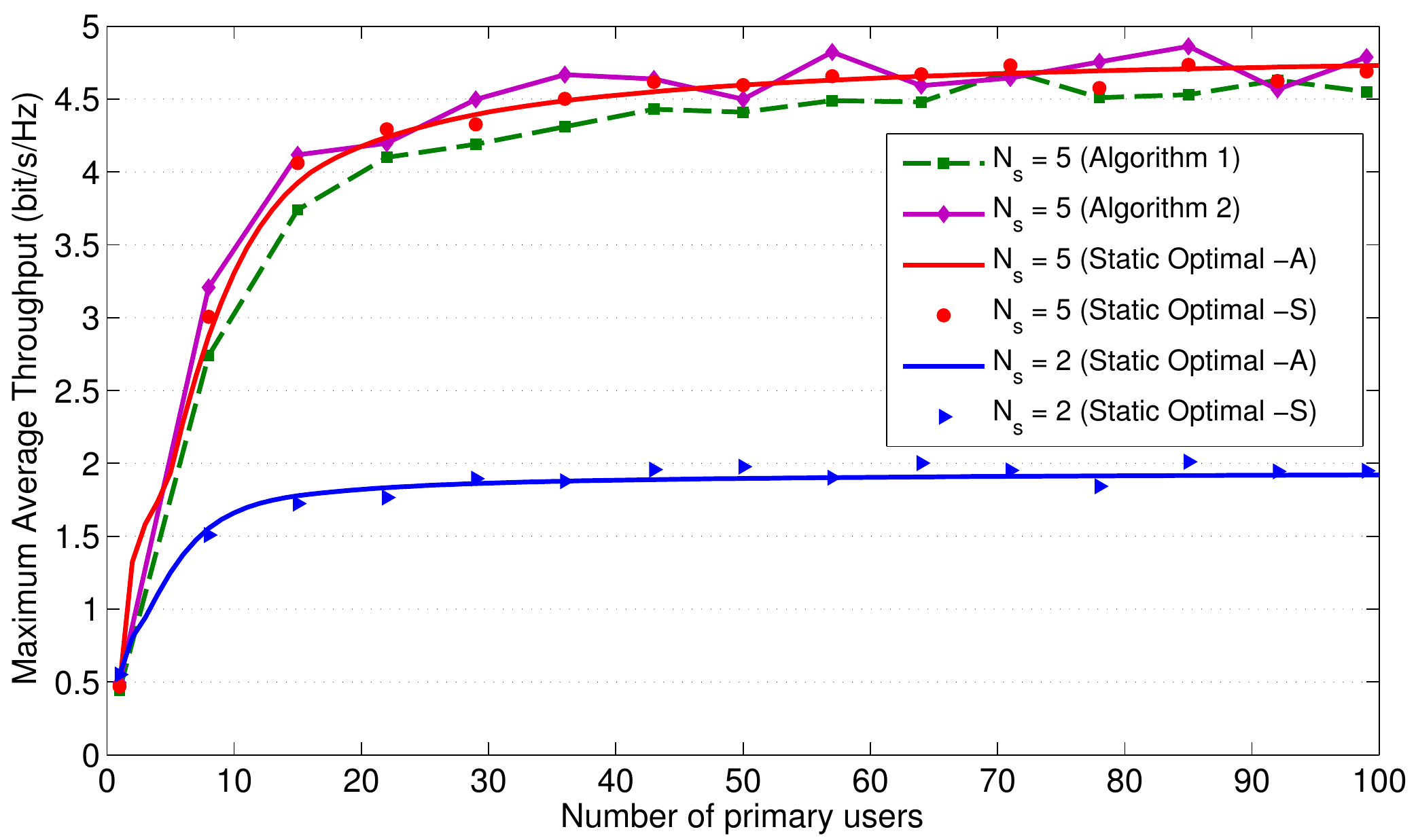}\\
  \caption{Average throughput against number of primary users. In the figure, -A stands for analysis and -S stands for simulations. Static optimal is the brute force solution of optimization problem \eqref{eq9}.}
  \label{fig: Max Throughput PUs}
\end{figure}

Fig.~\ref{fig: Max Throughput PUs} shows the maximum throughput of the network with respect to the number of primary channels. Three points can be made from the figure. First, mathematical derivations coincide numerical simulations, which further verifies our theoretical analysis.
Second, the maximum throughput raises with the number of primary channels in a saturating manner. This is due to that more channels are sensed, and therefore more transmission opportunities are found. Also, with extreme high number of primary channels, almost no collision happens among the SUs, and consequently a CRN with $N_s$ SUs can be modeled by $N_s$ distinct CRNs, each having one SU\footnote{According to the analyses provided in~\cite{shokriLB}, saturation of the maximum throughput for a CRN with one SU is expectable.}. For example, by looking at the maximum throughput for $N_p = 100$, both curves ($N_s = 2$ and $N_s = 5$) reach their saturating regions, and the maximum throughput when $5$ SUs exist in the CRN is around $2.5$ times of one achieved in the CRN with $2$ SUs.
Third, the proposed algorithms well mimics static optimal solution of \eqref{eq9}. More interestingly, Algorithm~\ref{Algorithm: Alg2} outperforms the static optimal throughput. The main reason is that all the SUs adopt similar values sensing time and also sensing probabilities, whereas Algorithm~\ref{Algorithm: Alg2} enables the SUs to adaptively adjust their sensing-access parameters in each sensing stage.
In fact, we have more degrees of freedom compared to static optimal design.

\begin{figure}[t]
\centering
  \includegraphics[width= 8.5cm]{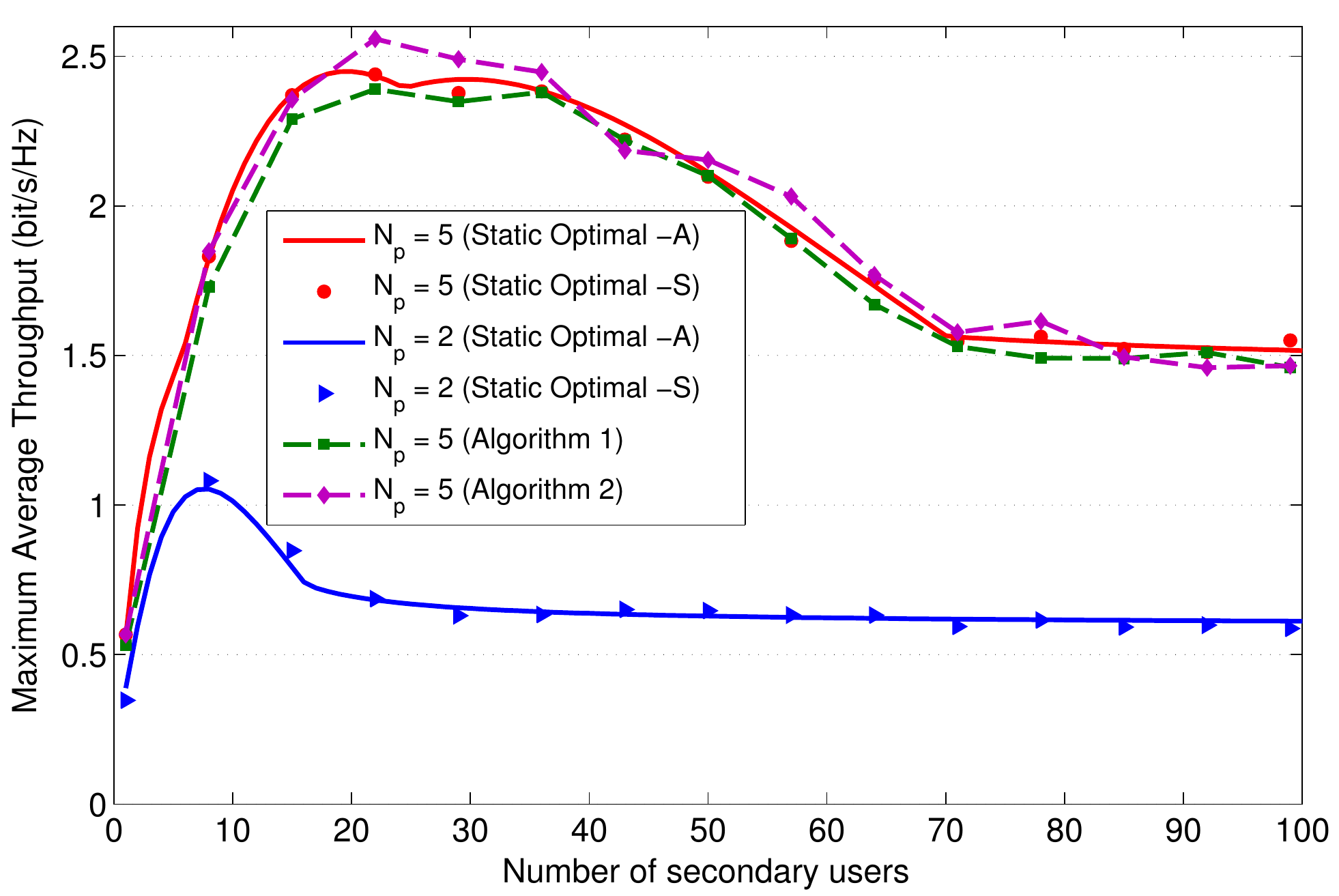}\\
  \caption{Average throughput against number of secondary users.}
  \label{fig: Max Throughput SUs}
\end{figure}

Fig.~\ref{fig: Max Throughput SUs} investigates the impact of secondary network size on the performance. The maximum throughput increases by the number of the SUs, but the contention level raises as well. Therefore, with more SUs in the network, each of them adopts lower value for $p$ to avoid collision with other SUs. this adjustment leads to less demand for sensing the channels and a high interest in being slept, which in turn results in wasting the transmission time. Clearly, this problem can be mitigated by increasing the number of primary channels, as can be observed in the figure. Again, the proposed algorithms perform well.
To more elaborate, Table \ref{table: performance comparison} demonstrates the performance enhancement due to optimal $p$ and $\tau$
derived in \eqref{eq9}, and compares the average throughput and interference for three different scenarios: (1) static optimal values, which are obtained by a brute force numerical optimization search and (2) adaptive values as achieved by the proposed Algorithm~\ref{Algorithm: Alg1}, (3) adaptive values as achieved by the proposed Algorithm~\ref{Algorithm: Alg2}. As expected, adopting the static optimal and adaptive values for $p$ and $\tau$ increases the average throughput while the interference meets the constraint.
Specifically, for the case $ N_s = 3, N_p = 7 $, the average throughput of
the SUs achieved by the static optimal design respectively is about $24\%$ and $2.1\%$ more than
the ones achieved in $p = 0.8, \tau = 0.1T$ (see Fig.~\ref{fig: Avg Thr and Int ver tau}) and adaptive algorithm.
Also, the average interference does not violate $t_I^{\max}$.

\begin{table}
\centering
  \caption{Average throughput (Thr.) and the corresponding normalized interference (Int.), for the static optimal parameters and adaptive algorithms 1 and 2.}
{\renewcommand{\arraystretch}{1.1}
    \begin{tabular}{c|c|c|c|c|c|c}
    \hhline{~------}
    \hhline{~------}
      & \multicolumn{2}{c|}{Optimal} & \multicolumn{2}{c|}{Algorithm 1} & \multicolumn{2}{c}{Algorithm 2} \\
    \hhline{~------}
      & Thr. & Int. & Thr. & Int. & Thr. & Int. \\
    \hline \hline
     $N_s = 3, N_p = 7$ & 1.441 & 0.047 & 1.411 & 0.042 & 1.447 & 0.039 \\
    \hline
     $N_s = 5, N_p = 7$ & 1.920 & 0.050 & 1.908 & 0.049  & 1.979 & 0.049 \\
    \hline
     $N_s = 7, N_p = 3$ & 1.371 & 0.050 & 1.340 & 0.047 & 1.366 & 0.043 \\
    \hline
    $N_s = 7, N_p = 5$ & 1.689 & 0.043 & 1.677 & 0.048 & 1.695 & 0.05 \\
    \hline
    \end{tabular}}
  \label{table: performance comparison}
\end{table}

\subsection{Optimality of Proposed Algorithms}
Although we have improved the current random sensing order policy for achieving better performance from throughput, interference, and energy perspectives, our approach actually gives an optimal solution only for the random sensing order policy. It might be that such a policy is not the optimal in general for other sensing order policies, however there are many compelling reasons for the specific policy we have considered. In particular, the adopted policy is a viable solution for a dense secondary network and the random access phase, where a centralized solution is not applicable.

\begin{figure}[t]
\centering
  \includegraphics[width= 8.5cm]{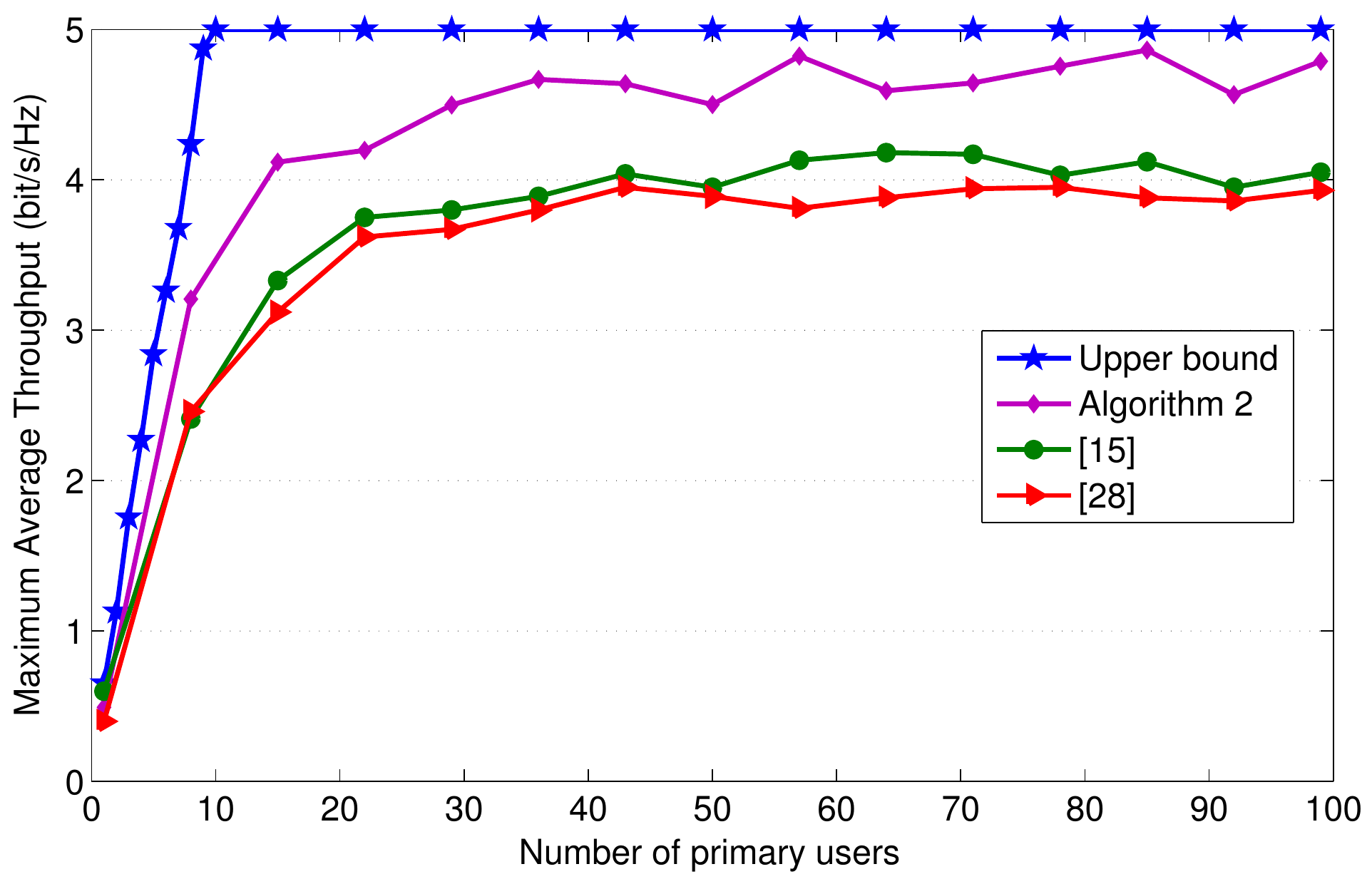}\\
  \caption{Average throughput against number of primary users. For Algorithm~\ref{Algorithm: Alg2}, we have duplicated ``$\rm N_s = 5$~(Algorithm~2)" from Fig.~\ref{fig: Max Throughput PUs}.}
  \label{fig: Comparison PUs}
\end{figure}

\begin{figure}[t]
\centering
  \includegraphics[width= 8.5cm]{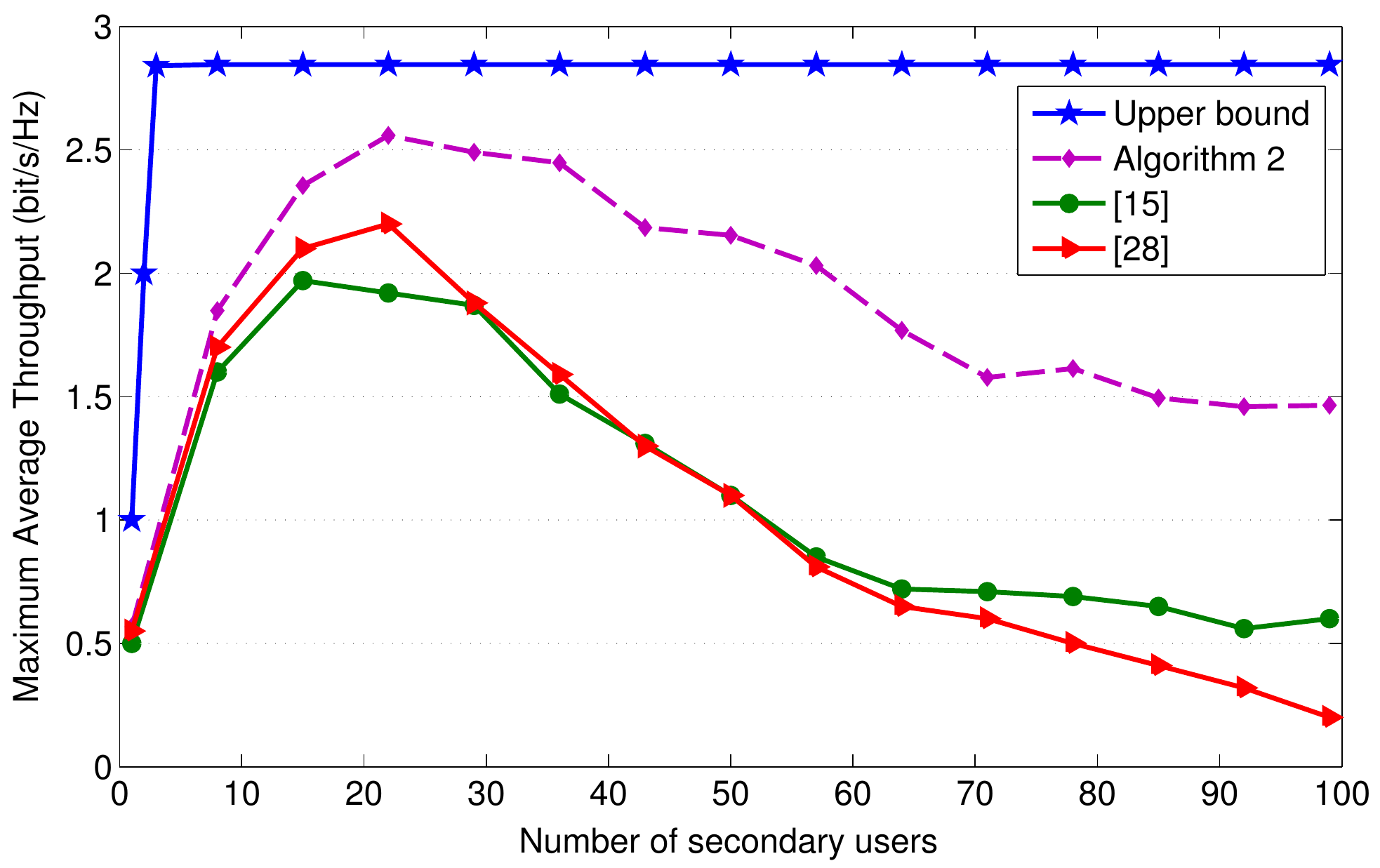}\\
  \caption{Average throughput against number of secondary users. For Algorithm~\ref{Algorithm: Alg2}, we have duplicated ``$\rm N_p = 5$~(Algorithm~2)" from Fig.~\ref{fig: Max Throughput SUs}.}
  \label{fig: Comparison_SUs}
\end{figure}

Figs.~\ref{fig: Comparison PUs} and~\ref{fig: Comparison_SUs} compares the performance of Algorithm~\ref{Algorithm: Alg2} to the algorithms introduced in~\cite{Li2013Almost} and~\cite{ShokriTWireless}. Also, we have added an ideal upper bound for the best achievable performance, which is obtained by assuming that all the available channels will be used by the SUs, without any false alarm, sensing time, interference, etc. This bound is dictated by two factors; namely the number of SUs and the average number of available primary channels. That is,
\begin{equation} \label{eq: UpperBound}
{\text{Upper bound}} = \min \left( N_s,\sum_{m=1}^{N_p}{\left(1 - P_{m,1} \right)} \right) \:,
\end{equation}
where $P_{m,1}$ is the presence probability of PU $m$, defined in Appendix A.

From Figs.~\ref{fig: Comparison PUs} and \ref{fig: Comparison_SUs}, we can draw the conclusion that the optimized random sensing order policy performs very well, even though it has not the best achievable performance, simply because we have considered almost the worst case scenario with the minimum signaling overheads. This result challenges the need of having complex solutions at least for non-dense secondary networks. For dense networks, for instance $N_s>50$ in Fig.~\ref{fig: Comparison_SUs}, the increased collision level of the network plays a dominant role in deterioration of the performance of the proposed algorithm, still it outperforms the distributed sensing order designs of~\cite{Li2013Almost} and~\cite{ShokriTWireless}.


\subsection{Energy Efficiency of Modified p-persistent Protocol}\label{section: modified vs conventional}
The advantages of our p-persistent protocol compared to the conventional one is illustrated in Table \ref{table: p-persistent comparison}. In the table, sensing overhead represents the average number of channels sensed by the SUs in a time slot. Considering a fixed energy consumption for sensing each channel~\cite{pei2011energy,Eryigit2013Energy}, the sensing overhead can be easily converted to energy consumption. As stated, the SUs always perform spectrum sensing in each stage, if conventional p-persistent random access algorithm is utilized~\cite{Wang2012Energ}. From the table, although both schemes can support multiple access among the SUs, but the modified scheme considered in this paper achieves the same average throughputs with considerable less consumed energies. For $ N_s = 7, N_p = 5 $, as an example, conventional protocol achieves $0.017 \%$ higher throughput at the expense of $17.47 \%$ more energy consumption compared to the modified protocol.
\begin{table}
\centering
  \caption{The performance comparison of the conventional and modified p-persistent schemes.}
{\renewcommand{\arraystretch}{1.1}
    \begin{tabular}{c|c|c|c|ccc}
    \hhline{~------}
    \hhline{~------}
      & \multicolumn{2}{c|}{Average throughput} & \multicolumn{2}{c}{Sensing overhead} \\
    \hhline{~------}
      & Conventional & Modified & Conventional & Modified\\
    \hline \hline
     $N_s = 3, N_p = 7$ & 1.6742 & 1.6746 & 4.5594 & 3.2422 \\
    \hline
     $N_s = 5, N_p = 7$ & 1.9349 & 1.9350 & 7.4922 & 5.9881 \\
    \hline
     $N_s = 7, N_p = 3$ & 1.2986 & 1.2981 & 9.6009 & 8.1273 \\
    \hline
    $N_s = 7, N_p = 5$ & 1.7655 & 1.7652 & 10.2214 & 8.7015 \\
    \hline
    \end{tabular}
  \label{table: p-persistent comparison}}
\end{table}

\section{Conclusion}\label{section: Conclusion}
Modeling and performance evaluation of random sensing order policy (RSOP) in a distributed cognitive radio network (CRN) were investigated in this paper. The behaviors of the secondary users (SUs) were modeled through a novel Markov process. The performance of the RSOP in terms of the average throughput of the CRN and average interference levels in the network was evaluated. Then, an optimization problem was formulated to maximize the average throughput while the interference level is kept bounded. Finally, to enhance the RSOP performance, two simple but efficient algorithms were proposed to adaptively adjust the sensing-access parameters. The convergence properties of the proposed algorithm were established. The algorithms enhance the performance of the CRN without high computational burden, as demonstrated through exhaustive numerical performance evaluations.

\section*{Appendix A} \label{section: Appendix A_New Occupancy}
Let $P_{m,1}$ denote the presence probability of PU $m$\footnote{ These
probabilities can be determined by modeling the PUs activities,
e.g., well-known ON-OFF model~\cite{wang2012modeling}.}. Also let $ P_{m,1}^{\left( n \right)}$ be occupation
probability of $m$-th channel at the beginning of $n$-th stage. At the beginning of the first stage, SUs have not sensed any channels yet, and therefore the occupation probability of each channel
is equal to the corresponding PU's presence probability. Thus, we
have
\begin{equation}\label{app_1}
P_{m,1}^{\left( 1 \right)} = P_{m,1} \qquad 1 \leq m \leq N_p \:.
\end{equation}

Let $\mathcal{N}_x$ be the number of the SUs that have requests
at the state $x$.
So, from Fig.~\ref{fig: MarkovChain}, we have, $\mathcal{N}_{{{\rm HO}_1}} = N_s$. The average number of
the SUs that sense the $m$-th channel at the first stage,
represented by $\mathcal{L}^{\left( 1 \right)}$, can be computed as
\begin{equation}\label{app_2}
\mathcal{L}^{\left( 1 \right)} = \mathcal{N}_{{{m}^{\left(1\right)}}} = \frac {p}{N_p} \mathcal{N}_{{{\rm HO}_1}} = p \frac {N_s}{N_p} \:.
\end{equation}

Each channel $m$ is sensed by $\mathcal{L}^{\left( 1
\right)}$ SUs at the first stage. Each of these SUs might sense
the corresponding channel free. In this case, the user starts its
transmission on the channel, and therefore contributes to this
channel's occupation probability. The probability of transmission on the $m$-th
channel by at least one SU conditioned on the absence of the PU is
\begin{equation}\label{app_3}
{U}_m ^{\left( 1 \right)} = 1 - { \left( P_{{\rm fa},m} ^{\left(1\right)} \right) ^{\mathcal{L}^{\left( 1 \right)}}} \:.
\end{equation}
${U}_m ^{\left( n \right)}$ is the probability that at least
one SU transmits on the $m$-th channel (or equivalently one SU
senses the channel free) at the end of the $n$-th stage conditioned on the absence of the PU.
Considering (\ref{app_1}), (\ref{app_2}), and (\ref{app_3}), we have
\begin{equation}\label{app_4}
P_{m,1}^{\left( 2 \right)} = {P_{m,1}^{\left( 1 \right)}} + {P_{m,0}}{{U}_m ^{\left( 1 \right)}} \:,
\end{equation}
and
\begin{align}\label{app_5}
\mathcal{N}_{{{\rm HO}_2}} &= {\left( 1 - p \right){\mathcal{N}_{{{\rm HO}_{1}}}} + \sum\limits_{m=1}^{N_p}{ {q_{m} ^{\left( 1 \right)}} {\mathcal{N}_{{m}^{\left(1\right)}}} }} \nonumber\\
&= {\left( \left( 1 - p \right) + {\frac{p}{N_p}}
\sum\limits_{m=1}^{N_p}{ {q_{m} ^{\left( 1 \right)}} }
 \right)}{\mathcal{N}_{{{\rm HO}_{1}}}} \:,
\end{align}
where $P_{m,0} = 1 - P_{m,1}$.

In Appendix B, it is proved that $P_{{\rm fa},m}^{\left( n \right)} = P_{{\rm fa},m}^{\left( 1 \right)}$ for $1 \leq m \leq N_p$ and $1 \leq n \leq \delta$. At the second stage, the number of SUs whose requests enter the state ${{{\rm HO}_2}}$ is calculated in (\ref{app_5}), where ${q_{m} ^{\left( n \right)}}$ is defined in \eqref{eq5}. We have
\begin{equation}\label{app_6}
{U}_m^{\left( 2 \right)} = 1 - {\left( P_{{\rm fa},m}^{\left( 2 \right)} \right) ^{\mathcal{L}^{\left( 2 \right)}}}
= 1 - {\left( P_{{\rm fa},m}^{\left( 1 \right)} \right) ^{\mathcal{L}^{\left( 2 \right)}}} \:,
\end{equation}
where $\mathcal{L}^{\left( 2 \right)} = \left( {p}/{N_p} \right) \mathcal{N}_{{{\rm HO}_2}} $. Therefore, the $m$-th channel occupation probability at the beginning of the third stage can be computed as
\begin{equation}\label{app_7}
\begin{split}
P_{m,1}^{\left( 3 \right)} &= P_{m,1} + {P_{m,0}}{{U}_m^{\left( 1 \right)}} + {P_{m,0}}{P_{{\rm fa},m} ^{\mathcal{L}^{\left( 1 \right)}}}{{U}_m ^{\left( 2 \right)}} \\
&= P_{m,1}^{\left( 2 \right)} + {P_{m,0}}{P_{{\rm fa},m} ^{\mathcal{L}^{\left( 1 \right)}}}{{U}_m ^{\left( 2 \right)}} \:.
\end{split}
\end{equation}
Following the same steps, at the $n$-th stage we have,
\begin{equation}\label{app_8}
P_{m,1}^{\left( n \right)} = P_{m,1}^{\left( n - 1 \right)} + {P_{m,0}}{ \left( P_{{\rm fa},m}^{\left( 1 \right)} \right) ^{\mathcal{L}^{\left( 1 \right)} + \mathcal{L}^{\left( 2 \right)} + \cdots + \mathcal{L}^{\left( n - 2\right)}}}{{U}_m ^{\left( n - 1 \right)}} \:,
\end{equation}
where
\begin{equation}\label{app_9}
{U}_m ^{\left( n \right)} = 1 - {\left( P_{{\rm fa},m} ^{\left( 1 \right)} \right) ^{\mathcal{L}^{\left( n \right)}}}
\end{equation}
\begin{equation}\label{app_10}
\mathcal{L}^{\left( n \right)} = \frac {p}{N_p} \mathcal{N}_{{{\rm HO}_n}} \:,
\end{equation}
and
\begin{equation}\label{app_11}
\mathcal{N}_{{{\rm HO}_n}} = {\left[ \left( 1 - p \right) + {\frac{p}{N_p}}
\sum\limits_{m=1}^{N_p}{ {q_{m} ^{\left( n - 1 \right)}} }
 \right]}{\mathcal{N}_{{{\rm HO}_{n - 1}}}} \:.
\tag*{$\blacksquare$}
\end{equation}

\section*{Appendix B} \label{section: Appendix B New Pmd}
The state of channel $k$, which is dedicated to $k$-th PU, is represented by
\begin{equation}\label{eq201}
s_k \left( t \right ) = \left\{ \begin{matrix}
1 ~ :& \quad {\mathcal{H}}_1 \\
0 ~ :& \quad {\mathcal{H}}_0
\end{matrix}\right. \:,
\end{equation}
where $\mathcal{H}_0$ and $\mathcal{H}_1$ respectively denote the
occupancy and idleness hypotheses of the channel $k$.
Recall that each channel may be occupied by the corresponding PU or/and some SUs.
Here, for simplicity of presentation, we only formulate false alarm and misdetection probabilities for an AWGN channel. The formulations, however, can easily be extended to consider more realistic channels~\cite{Digham2003On}.
For an AWGN channel, the spectrum sensing process is modeled as a binary
hypotheses testing problem~\cite{liang}
\begin{equation}\label{eq202}
\left\{ \begin{matrix}
\mathcal{H}_0~: & y \left( k \right) = z \left( k \right) \\
\mathcal{H}_1~: & y \left( k \right) = u_m \left( k \right) + z \left( k \right)
\end{matrix}\right. \:,
\end{equation}
where $z \left( k \right)$ is $k$-th sample of zero mean complex-valued Gaussian noise with independent and identical
distribution (i.i.d). $u_m \left( k \right)$ and $y \left( k \right)$ denote the $k$-th sample of the accumulated signal (exclude noise) that presents in the channel $m$, which is independent of $z \left( k \right)$, and the $k$-th sample of the received signal.

There are various spectrum sensing proposals~\cite{Arsalan09}. Among them, we derive the formulations for energy detector; because it is the most prevalent spectrum sensing scheme in the literature. Also, it is the optimal detector for unknown signals~\cite{Arsalan09}. To decide about occupation status of a channel, in the energy detector scheme, energy of a received signal is accumulated during a sensing time $\tau$, and then it is compared to a threshold $\lambda$. Let $w = {\tau}{f_s}$ represent the number of samples taken from the received signal, where $f_s$ is the sampling frequency.
Defining $X$ as the accumulated energy of $w$ consecutive samples, the decision criteria is defined as
\begin{equation}\label{eq203}
X = \sum_{k=1}^{w}{{\left | y \left ( k \right ) \right |}^2} =
\left\{ \begin{matrix}
< \lambda ~: & \mathcal{H}_0 \\
\geq \lambda ~: & \mathcal{H}_1
\end{matrix}\right. \:.
\end{equation}
Considering a gaussian distribution for $X$ (which is meaningful for large $w$~\cite{liang}), we have~\cite{akyildiz2010cooperative}
\begin{equation}\label{eq204}
P_{{\rm fa},m}^{\left( n \right)} = \mathcal{Q} \left( \left( \frac{\lambda}{\sigma _{z}
^2} - 1 \right) \sqrt{\tau f_s} \right) \:,
\end{equation}
\begin{equation}\label{eq205}
P_{{\rm md},m}^{\left( n \right)} = 1 - \mathcal{Q} \left( \left( \frac{\lambda}{\sigma
_{z} ^2} - 1 - \gamma_m^{\left( n \right)} \right) \sqrt{\frac{\tau f_s}{1 + 2
\gamma_m^{\left( n \right)}}} \right) \:,
\end{equation}
where $P_{{\rm fa},m}^{\left( n \right)}$ and $P_{{\rm md},m}^{\left( n \right)}$ respectively are the false alarm and misdetection probabilities when an SU senses the channel $m$ at stage $n$.
${\sigma_{z} ^2}$ is the noise variance, and $\gamma_m^{\left( n \right)}$ is the received signal to the noise ratio of the channel $m$ at the stage $n$.
Suppose that $\sigma_{p_m}^{\left( 2 \right)}$ is the power of the PU $m$ at the secondary receiver. Let $\sigma_s^{\left( 2 \right)}$ be the power of each SU.
At the beginning of each time slot, the occupancy status of the channels only depends the PUs activities. Hence,
\begin{equation}\label{eq206}
\gamma_{m}^{\left( 1 \right)} = \frac{\sigma_{p_m}^{\left( 2 \right)}}{\sigma_{z}^{\left( 2 \right)}} \:.
\end{equation}
Also, considering the definition of $\mathcal{L}^{\left( 1 \right)}$, introduced in \eqref{app_2}, $p{N_s/N_p} \left( 1 - q_m^{\left( 1 \right)} \right)$ SUs transmit on the channel $m$ at the first stage. Therefore, the remaind SUs take samples from a received signal with an SNR
\begin{equation}\label{eq207}
\gamma_{m}^{\left( 2 \right)} = \frac{P_{m,1} \sigma_{p_m}^{\left( 2 \right)} + {N_s}{p/N_p} \left( 1 - q_m^{\left( 1 \right)} \right) \sigma_{s}^{\left( 2 \right)}}{\sigma_{z}^{\left( 2 \right)}} \:,
\end{equation}
when they intend to sense the channel $m$ at the second stage. $P_{m,1}$ is as defined in \eqref{app_1}.

A false alarm occurs when a free channel is mistakenly sensed busy. Consequently, there are no PU or SUs signals on the channel when a false alarm happens. Therefore, the possible changes in the level of the signals in successive stages do not affect the false alarm probability, as can be concluded from \eqref{eq204}.
For each channel $m$ and each stage $n$, $P_{{\rm fa},m}^{\left( n \right)} = P_{{\rm fa},m}^{\left( 1 \right)}$ for $1\leq m \leq N_p$.
On the other hand, the misdetection probability directly relates to the received SNR and may change in different stages, depending on the number of SUs transmit on the corresponding channel. $P_{d,m}^{\left( 1 \right)}$ and $P_{d,m}^{\left( 2 \right)}$ can be computed by substituting \eqref{eq206} and \eqref{eq207} into \eqref{eq205}.
\begin{table}
  \centering
  \caption{Detection probability in sensing stages}\label{table: detection_probability}
{\renewcommand{\arraystretch}{1.1}
   \begin{tabular}{|c |c |c |c |c |c | c|} \hline
   Stage 1 & Stage 2 & Stage 3 & Stage 4 & Stage 5 & Stage 6 & Stage 7 \\ \hline
    0.933  &  0.986  &  0.992  &  0.994  &  0.995  &  0.996  &  0.997  \\
  \hline
\end{tabular}}
\end{table}

The increment of the detection probability has a saturating pattern regarding the SNR value~\cite{Ma2008Soft}. That is, a small enhancement in the SNR manifests itself as a significant enhancement in the detection performance, and then further increment of the SNR value does not improve the detection performance substantially. From \eqref{eq207}, the SNR of the channel $m$ is significantly increased in the second stage in the average sense, leading to a meaningful improvement in the detection probability. However, we expect that the detection performance will not be dramatically enhanced in the next sensing stages due to the aforementioned saturating pattern. To more illustrate this phenomenon, in the following we consider a numerical example.
Table~\ref{table: detection_probability} shows the detection probability for consecutive sensing stages, assuming $T = 10 ~{\text{ms}}$, $\tau = 0.1T$, $\tau_{h} = 0.1 \mu ~{\text{s}}$, $\sigma_{p_m}^{\left( 2 \right)} = \sigma_{s}^{\left( 2 \right)}$, $N_s = 20$, $N_p = 10$, and $p = 0.8$. The simulation setup procedure is described in Section \ref{section: Numerical Results}. From the table, the detection probability is not substantially improved after the second sensing stage. Clearly, the enhancement ratio highly depends on the number of SUs and PUs, and also the channel sensing probability $p$. Although the SNR in the second stage is higher than the SNR of the first stage regardless of the exact values of the above parameters, the small enhancement in the SNR can drive the detection probability to its saturation region. Here, we assume that
\begin{equation}\label{eq208}
P_{{\rm md},m}^{\left( n \right)} \approx P_{{\rm md},m}^{\left( 2 \right)} \qquad 3 \leq n \leq \delta \:.
\tag*{$\blacksquare$}
\end{equation}

\section*{Appendix C} \label{section: Appendix C_Performance Evaluation}
Let ${\Pi _x}$ denote the probability of being at
state $x$. From Fig.~\ref{fig: MarkovChain}, we have
\begin{equation}\label{Ap301}
\Pi _{{\rm HO}_{n}} = {\left[ \left( 1 - p \right) + {\frac{p}{N_p}}
\sum\limits_{m=1}^{N_p}{ {q_{m} ^{\left( n - 1 \right)}} }
 \right]}\Pi _{{\rm HO}_{n-1}} \:,
\end{equation}
\begin{equation}\label{Ap302}
\Pi _{{{m}}^{\left( n \right)}} = \frac{p}{N_p} \Pi _{{\rm HO}_{\left( n \right)}} \:,
\end{equation}
\begin{equation}\label{Ap303}
\Pi _{{I}_{n}} = \sum\limits_{m=1}^{N_p} {P_{m,1}^{\left( n \right)}\left( {1 - {P_{d,m}^{\left( n \right)}}} \right) \Pi _{{{m}}^{\left( n \right)}}} \:,
\end{equation}
and
\begin{equation}\label{Ap304}
\Pi _{{T}_{n}} = \sum\limits_{m=1}^{N_p} {P_{m,0}^{\left( n \right)}\left( {1 - {P_{{\rm fa},m}^{\left( n \right)}}} \right) \Pi _{{{m}}^{\left( n \right)}}} \:.
\end{equation}
Note that considering the channel search and access policy described in Section \ref{section: RSOP}, the procedure always initiates from the state ${\rm HO}_1$, and thus
\begin{equation}\label{Ap310}
\Pi _{{\rm HO}_1} = 1 \:.
\end{equation}
Then, from \eqref{Ap301}--\eqref{Ap304}, $\Pi _{{I}_{n}}$ and $\Pi _{{T}_{n}}$ are calculated.
Let $P_{T_n,{m}^{\left( n \right)}}^{\left[ k \right]}$ and $P_{I_n,{m}^{\left( n \right)}}^{\left[ k \right]}$ respectively denote the probability of the transmissions of the SU $k$ from the states $T_n$ and $I_n$ on the channel $m$, i.e., the state changes from the state ${m}^{\left( n \right)}$ to the states $T_n$ and $I_n$:
\begin{equation}\label{Ap311}
P_{T_n,{m}^{\left( n \right)}}^{\left[ k \right]} = {{\Pi _{{{{m}}^{\left( n \right)}}}^{\left[ k \right]}}} {P_{m,0}^{\left( n \right)}\left( {1 - {P_{{\rm fa},m}^{\left( n \right)}}} \right) }
\end{equation}
\begin{equation}\label{Ap312}
P_{I_n,{m}^{\left( n \right)}}^{\left[ k \right]} = {{\Pi _{{{{m}}^{\left( n \right)}}}^{\left[ k \right]}}} {P_{m,1}^{\left( n \right)}\left( {1 - {P_{d,m}^{\left( n \right)}}} \right) } \:.
\end{equation}

The $k$-th SU will successfully transmit data on each channel $m$ at the stage $n$ (with probability ${Q}_{T_n,m}^{\left[ k \right]}$) provided that its state transits from states ${m}^{\left( n \right)}$ to $T_n$ for $1 \leq n \leq \delta$ (with probability $P_{T_n,{m}^{\left( n \right)}}^{\left[ k \right]}$), and all other SUs do not collide its communications. Assume that ${Y}_{m,n}^{\left[ \ell \right]}$ represents the probability of the SU $\ell$ does not transmit on the channel $m$ in stages $n,n+1,\ldots,\delta$. Hence, the $k$-th SU successfully transmits data on the channel $m$ at the stage $n$ with probability
\begin{equation}\label{Ap313}
{Q}_{T_n,m}^{\left[ k \right]} = {\prod\limits_{\scriptstyle {\ell = 1} \hfill \atop
  \scriptstyle \ell \ne k \hfill} ^{{N_s}} {P_{T_n,{m}^{\left( n \right)}}^{\left[ k \right]}{Y}_{m,n}^{\left[ \ell \right]}}} \:.
\end{equation}
If we omit the superscript $\left[ k \right]$, \eqref{Ap313} is simplified to ${Q}_{T_n,m} = P_{T_n,{m}^{\left( n \right)}}{Y}_{m,n}^{N_s - 1}$. At the stage, the SU transmits for ${\rm{RT}}_n$ time units with the constant rate of $C_R$. The average throughput of each SU follows as
\begin{equation}\label{Ap314}
r = \frac{1}{T} \sum\limits_{{m}=1}^{N_p} {\sum\limits_{n=1}^{\delta} {{Q}_{T_n,m} {\rm{RT}}_n C_R}} \:.
\end{equation}

\begin{figure}[t]
\centering
  \includegraphics[width= 8cm]{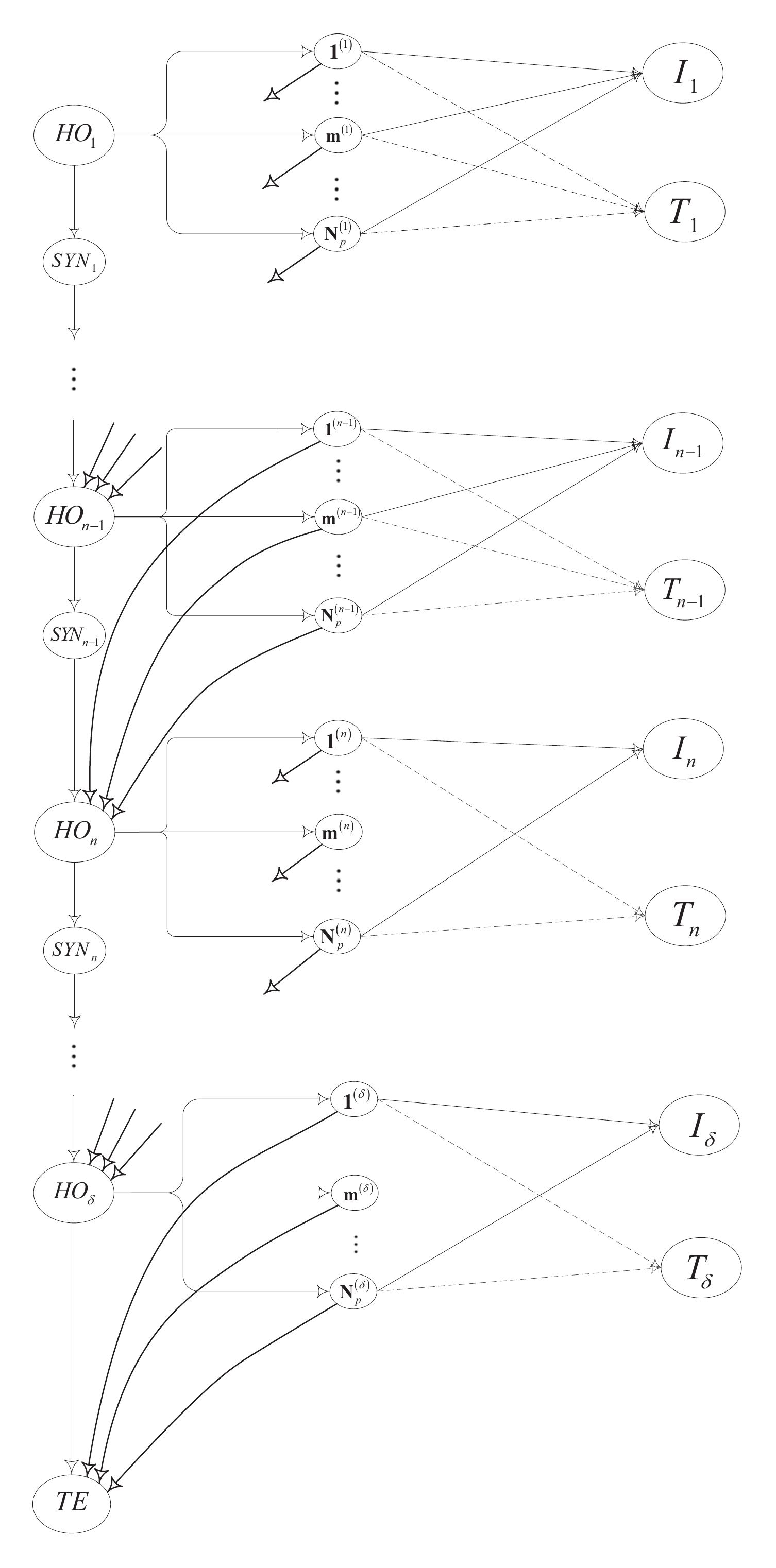}\\
  \caption{Pruned Markov structure to model the event, wherein the SU $k$ does not transmit on the channel $m$ in stages $n,n+1,\ldots,\delta$. Labels of the transition probabilities have been removed to make the figure more clear.}
  \label{fig: Pruned MarkovChain}
\end{figure}

Finally, we need to formulate ${Y}_{m,n}$. Assume that an SU starts transmission on each channel $m$ at each stage $n$. In this case, if another SU selects the same channel at the same stage, i.e., the channel $m$ at the stage $n$, it must go to the next handoff state. It is equal to remove the edges between state $\mathbf{m}^{\left( n \right)}$ and $T_n$ and $I_n$ from Fig.~\ref{fig: MarkovChain}. Moreover, all the edges between state $\mathbf{m}^{\left( j \right)}$ and $T_j$ and $I_j$, $n+1 \leq j \leq \delta$, of the remaining $N-1$ SUs must be removed to avoid any possible interference with an ongoing SU transmission, which is initiated in previous stages. Altogether, in the pruned Markov model, we have the same structure as Fig.~\ref{fig: MarkovChain} for the stages $1,2,\ldots,n-1$, however all the edges between state $\mathbf{m}^{\left( n \right)}$ and $T_n$ and $I_n$ for $n, n+1, \ldots,\delta$ are removed.
Fig.~\ref{fig: Pruned MarkovChain} depicts the pruned Markov model. Using this figure and following the same steps taken in the Appendices A and B, the probabilities of being at $T_n$, $I_n$ or eventually ${\rm TE}$ is obtained. Then, we have
\begin{equation}\label{Ap315}
{Y}_{m,n} = \Pi_{{\rm TE}} + \sum\limits_{i=1}^{\delta} {\Pi_{T_i} + \Pi_{I_i}} \:.
\end{equation}

To find the average interference time, $t_I$, note that each SU encounters
\begin{equation*}
t_I^{\left[ k \right]} = \frac{1}{TN_p} \sum\limits_{n=1}^{\delta} {\Pi _{I_n} {\rm{RT}}_n}
\end{equation*}
level of interference in each time slot. But, these random variables are not independent; because SU $k_2$ can transmit on the occupied channel $m$, where it has been mistakenly interfered by SU $k_1$ in the previous sensing stages. Therefore,
\begin{equation*}
\frac{1}{TN_p} \sum\limits_{k=1}^{N_s} {\sum\limits_{n=1}^{\delta} {\Pi _{I_n} {\rm{RT}}_n}}
\end{equation*}
is an upper bound of the interference time of the network. Let ${Z}_{I_n,m}$ be the probability no SUs cause interference on the channel $m$ at the stage $n$. From \eqref{Ap312}, we have
\begin{equation}\label{Ap316}
{Z}_{I_n,m} = \prod\limits_{k = 1} ^{N_s} {\left (1 - P_{I_n,{m}^{\left( n \right)}}^{\left[ k \right]}\right)} = \left( 1 - P_{I_n,{m}^{\left( n \right)}} \right)^{N_s} \:.
\end{equation}
Hence, the average interference time of the network is:
\begin{equation}\label{Ap317}
t_I = \frac{1}{T N_p} {\sum\limits_{m = 1} ^{N_p} {\sum\limits_{n = 1} ^{\delta} { \left( 1 - {Z}_{I_n,m} \right) {\rm{RT}}_n}}} \:.
\end{equation}

Note that we have to neglect minor impacts of parameters on each others to obtain analytic relationships. This is a common assumption that is widely adopted, for instance in~\cite{park2009generalized}, following the well-known Bianchi's Markov chain model~\cite{bianchi2000performance} and without which the analysis would be a formidable task, if not impossible. \hspace*{\fill}{$\blacksquare$}

\section*{Appendix D} \label{section: Appendix D_proof_subgradient}

\begin{figure}
\centering
\hspace{-1cm}
  \includegraphics[width= 8.5cm]{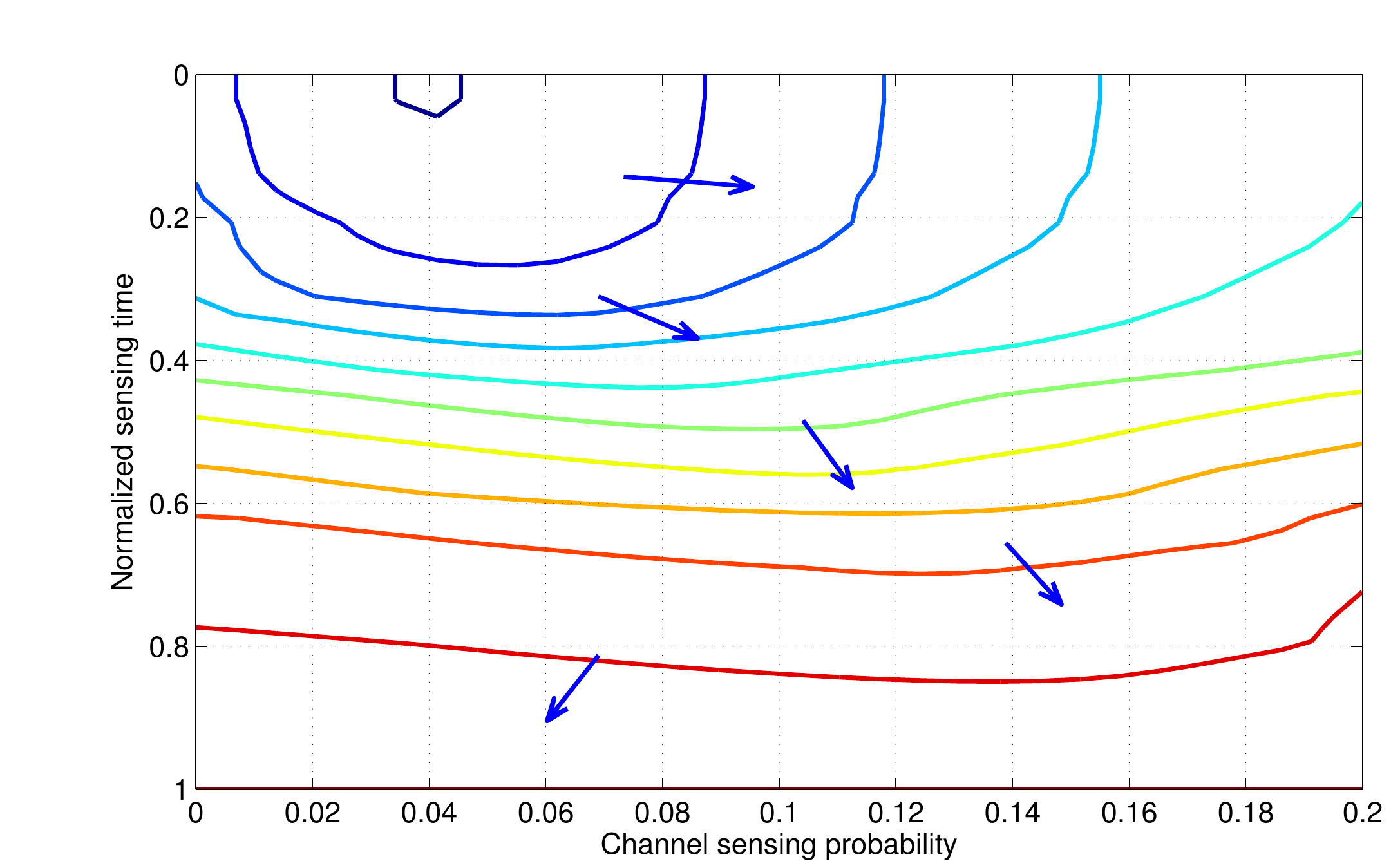}\\
  \caption{Subgradient of the objective function, for $N_s = 5$ and $N_p = 5$.}
  \label{fig: Subgradient}
\end{figure}

Due to formidable complexity of an analytical investigation, we have used extensive Monte Carlo simulations to show that ${\mathbb{E}[\widetilde{\mathbf{g}}^{k} | {\mathbf{x}^{k}}]}$ belongs the set of subgradients of the objective function at ${\mathbf{x}^{k}}$ for feasible points. This simple procedure, i.e., Monte Carlo Simulations, is common in the optimization literature when showing properties of gradients and subgradients constitutes a formidable or impossible task~\cite{boyd2004convex}.
To this end, for a given feasible ${\mathbf{x}^{k}}$, we have run Algorithm 1, 5000 times. In each realization, the average throughput and interference time, thus $\widetilde{\mathbf{g}}_{m}^{k}$ have been estimated for each SU $m$. To make comparison possible, we have computed the average $\widetilde{\mathbf{g}}_{m}^{k}$ over all SUs as the estimated subgradient at ${\mathbf{x}^{k}}$ for this realization. Then, we have calculated ${\mathbb{E}[\widetilde{\mathbf{g}}^{k} | {\mathbf{x}^{k}}]}$ by averaging over 5000 realizations. Fig.~\ref{fig: Subgradient} shows contours of the objective function for $N_s = 5$ and $N_p = 5$. Each arrow represents ${\mathbb{E}[\widetilde{\mathbf{g}}^{k} | {\mathbf{x}^{k}}]}$ that has been perfectly aligned to the gradient of the objective function at ${\mathbf{x}^{k}}$, computed numerically. Please note that we have observed identical behavior for different $N_s$ and $N_p$, confirming the above claim for all cases. Such behaviors are also expected since the algorithm always converges to the optimal point, as shown in Figures 7 and 8 and Table V of the manuscript.
\bibliographystyle{IEEEtran}
\bibliography{IEEEabrv,biblio}

\begin{IEEEbiography}[{\includegraphics[width=1in,height=1.25in]{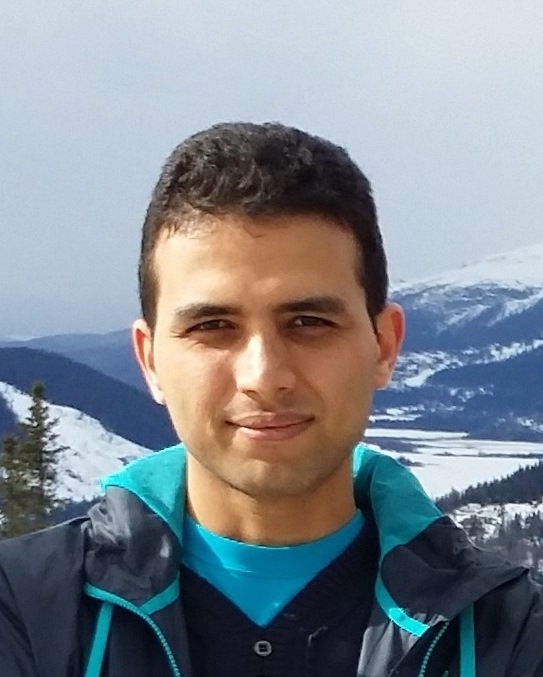}}]%
{Hossein Shokri-Ghadikolaei} received the B.S. and M.S. degrees in communication systems from Iran University of Science and Technology and Sharif University of Technology, Tehran, Iran, in 2009 and 2011, respectively.
He is currently working toward the Ph.D. degree in the School of Electrical Engineering, KTH Royal Institute of Technology, Stockholm, Sweden.
His research interests include performance analysis and optimization of directional communications, with application in millimeter wave communications, and cognitive networks.

He received a number of awards, including the best paper award from the Iranian Student Conference of Electrical Engineering, 2011, IEEE ComSoc top ten paper, 2013, Program of Excellence award from KTH, 2013, and Premium Award for Best Paper in IET Communications, 2014.
He is a member of working group 1900.1 in the IEEE Dynamic Spectrum Access Networks Standards Committee (DySPAN-SC).
He has chaired or served as a technical member of program committees of several international conferences and is serving as referee for technical journals.
\end{IEEEbiography}

\vspace{-5 cm}

\begin{IEEEbiography}[{\includegraphics[width=1in,height=1.25in,keepaspectratio]{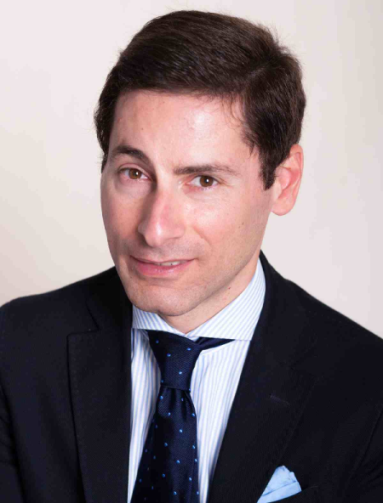}}]%
{Carlo Fischione} is a tenured Associate Professor with KTH Royal Institute of Technology, Electrical Engineering and ACCESS Linnaeus Center, Automatic Control Lab, Stockholm, Sweden. He received the Dr.Eng. degree in Electronic Engineering and the Ph.D. degree in Electrical and Information Engineering from the University of L'Aquila, L'Aquila, Italy, in 2001 and 2005, respectively.
He held research positions with University of California at Berkeley, Berkeley, CA (2004-2005, Visiting Scholar, and 2007-2008, Research Associate) and Royal Institute of Technology, Stockholm, Sweden (2005-2007, Research Associate). His research interests include optimization and parallel computation with applications in wireless sensor networks, networked control systems, and wireless networks. He has co-authored over 80 publications, including book, book chapters, international journals and conferences, and an international patent.

He received a number of awards, including the best paper award from the IEEE Transactions on Industrial Informatics of 2007, the best paper awards at the IEEE International Conference on Mobile Ad-hoc and Sensor System 05 and 09 (IEEE MASS 2005 and IEEE MASS 2009), the Best Business Idea award from VentureCup East Sweden, 2010, the “Ferdinando Filauro” award from University of L'Aquila, Italy, 2003, the “Higher Education” award from Abruzzo Region Government, Italy, 2004, the Junior Research award from Swedish Research Council, 2007, and the Silver Ear of Wheat award in history from the Municipality of Tornimparte, Italy, 2012. He has chaired or served as a technical member of program committees of several international conferences and is serving as referee for technical journals. Meanwhile, he also has offered his advice as a consultant to numerous technology companies such as Berkeley Wireless Sensor Network Lab, Ericsson Research, Synopsys, and United Technology Research Center. He is co-funder and CTO of the sensor networks start-up company Aukoti. He is Member of IEEE (the Institute of Electrical and Electronic Engineers), and Ordinary Member of DASP (the academy of history Deputazione Abruzzese di Storia Patria).
\end{IEEEbiography}

\end{document}